\documentclass{article}%
\usepackage{graphicx}
\usepackage{amsmath}
\usepackage{amsfonts}
\usepackage{amssymb}
\usepackage[doublespacing]{setspace}%
\newtheorem{theorem}{Theorem}
{}
\newtheorem{acknowledgement}{Acknowledgement}

\newtheorem{corollary}{Corollary}

\newtheorem{lemma}{Lemma}
{}

\newtheorem{proposition}{Proposition}
\newtheorem{remark}{Remark}

\newenvironment{proof}[1][Proof]{\textbf{#1.} }{\ \rule{0.5em}{0.5em}}

\begin{document}

\author{O. A. Veliev\\{\small \ Department of Mathematics, Dogus University, Acibadem, }\\{\small Kadikoy, Istanbul, Turkey, e-mail: oveliev@dogus.edu.tr}}
\title{An Algorithm for Finding the Periodic Potential of the Three-Dimensional
Schr\"{o}dinger Operator from the Spectral Invariants }
\date{}
\maketitle

\begin{abstract}
In this paper, we investigate the three-dimensional Schr\"{o}dinger operator
with a periodic, relative to a lattice $\Omega$ of $\mathbb{R}^{3},$ potential
$q.$ A special class $V$ of the periodic potentials is constructed, which is
easily and constructively determined from the spectral invariants. First, we
give an algorithm for the unique determination of the potential $q\in V$ of
the three-dimensional Schr\"{o}dinger operator from the spectral invariants
that were determined constructively from the given Bloch eigenvalues. Then we
consider the stability of the algorithm with respect to the spectral
invariants and Bloch eigenvalues. Finally, we prove that there are no other
periodic potentials in the set of large class of functions whose Bloch
eigenvalues coincides with the Bloch eigenvalues of $q\in V.$

Keywords: Schr\"{o}dinger operator, spectral invariants, inverse problem.

Mathematics Subject Classifications : 47F05, 35J10

\end{abstract}

\section{Introduction}

Let $L(q)$ be the Schr\"{o}dinger operator%
\begin{equation}
L(q)=-\Delta+q(x),\ x\in\mathbb{R}^{d}\
\end{equation}
with a periodic, relative to a lattice $\Omega$, potential $q(x).$ The
operator $L(q)$ describes the motion of a particle in bulk matter. Therefore,
for physical applications, it is important to have a detailed analysis of the
spectral properties of $L(q)$. Let $F=:\mathbb{R}^{d}/\Omega$ be a fundamental
domain of $\Omega$ and $L_{t}(q)$ be the operator generated in $L_{2}(F)$ by
(1) and the conditions:%
\[
u(x+\omega)=e^{i\left\langle t,\omega\right\rangle }u(x),\ \forall\omega
\in\Omega,
\]
where $t\in F^{\star}=:\mathbb{R}^{d}/\Gamma,$ $\Gamma$ is the lattice dual to
$\Omega$, that is, $\Gamma$ is the set of all vectors $\gamma\in\mathbb{R}%
^{d}$ satisfying $\langle\gamma,\omega\rangle\in2\pi\mathbb{Z}$ for all
$\omega\in\Omega,$ and $\langle.,.\rangle$ is the inner product in
$\mathbb{R}^{d}.$ It is well-known that (see [1])\ the spectrum of $L(q)$ is
the union of the spectra of $L_{t}(q)$ for $t\in F^{\ast}$. The eigenvalues
$\Lambda_{1}(t)\leq\Lambda_{2}(t)\leq...$ of $L_{t}(q)$ are called the Bloch
eigenvalues of $\ L(q).$ These eigenvalues define the continuous functions
$\Lambda_{1}(t),$ $\Lambda_{2}(t),...$ of $t$ that are called the band
functions of $L(q)$. The intervals $\{\Lambda_{1}(t):t\in F^{\star}\},$
$\{\Lambda_{2}(t):t\in F^{\star}\},...$ are the bands of the spectrum of
$L(q)$ and the spaces (if exists) $\Delta_{1},\Delta_{2},...$ between two
neighboring bands are the gaps in the spectrum. Without loss of generality, we
assume that the measure of $F$ is $1.$

In this paper we determine constructively and uniquely, modulo the following
inversion and translations
\begin{equation}
q(x)\rightarrow q(-x),\ q(x)\rightarrow q(x+\tau),\ \tau\in\mathbb{R}^{d},
\end{equation}
the potential $q(x)$ of the three- dimensional ($d=3$) Schr\"{o}dinder
operator $L(q)$ from the given spectral invariants , when $q(x)$ has the form
\begin{equation}
q(x)=%
%TCIMACRO{\tsum \limits_{a\in Q(1,1,1)}}%
%BeginExpansion
{\textstyle\sum\limits_{a\in Q(1,1,1)}}
%EndExpansion
z(a)e^{i\langle a,x\rangle},
\end{equation}
where%
\begin{equation}
z(a)=:(q(x),e^{i\langle a,x\rangle})\neq0,\text{ }\forall a\in Q(1,1,1),
\end{equation}%
\begin{equation}
Q(1,1,1)=:\{n\gamma_{1}+m\gamma_{2}+s\gamma_{3}:|n|\leq1,\ |m|\leq
1,\ |s|\leq1\}\backslash\{(0,0,0)\},
\end{equation}
$(.,.)$ is the inner product in $L_{2}(F)$ and $\{\gamma_{1},\gamma_{2}%
,\gamma_{3}\}$ is a basis of $\Gamma$ satisfying
\begin{equation}
\langle\gamma_{i},\gamma_{j}\rangle\neq0,\text{ }\langle\gamma_{i}+\gamma
_{j},\gamma_{k}\rangle\neq0,\text{ }\mid\gamma_{i}\mid\neq\mid\gamma_{j}%
\mid,\text{ }\langle\gamma_{i}+\gamma_{j}+\gamma_{k},\gamma_{i}-\gamma
_{j}-\gamma_{k}\rangle\neq0
\end{equation}
for all different indices $i,j,k$. Note that every lattice has a basis
satisfying (6) (see Proposition 2 in Section 2) and the potential $q$ can be
uniquely determined only by fixing the inversion and translations (2), since
the operators $L(q(x)),\ L(q(-x)),$ $L(q(x+\tau))$ have the same band functions.

The inverse problem of the one-dimensional Schr\"{o}dinger operator, that is,
the Hill operator, denoted by $H(q)$, and the multidimensional Schr\"{o}dinger
operator $L(q)$ are absolutely different. Inverse spectral theory for the Hill
operator has a long history and there exist many books and papers about it
(see, for example, [6] and [7]). In order to determine the potential $q,$
where $q(x+\pi)=q(x),$ of the Hill operator, in addition to the given band
functions $\Lambda_{1}(t),$ $\Lambda_{2}(t),...$ , one needs to know the
eigenvalues $\lambda_{1},\lambda_{2},...$ of the Dirichlet boundary value
problem and the signs of the numbers $u_{-}(\sqrt{\lambda_{1}}),$ $u_{-}%
(\sqrt{\lambda_{2}}),...$ , where $u_{-}(\lambda)=c(\lambda,\pi)-s^{^{\prime}%
}(\lambda,\pi)$ and $c(\lambda,x),$ $s(\lambda,x)$ are the solutions of the
Hill equation%
\[
-y^{^{\prime\prime}}(x)+q(x)y(x)=\lambda^{2}y(x)
\]
satisfying $c(\lambda,0)=s^{^{\prime}}(\lambda,0)=1,$ $c^{^{\prime}}%
(\lambda,0)=s(\lambda,0)=0$ (see [7], Chap.3, Sec. 4). In other words, the
potential $q$ of the Hill operator can not be determined uniquely from the
given band functions, since if the band functions $\Lambda_{1}(t),$
$\Lambda_{2}(t),...$ of $H(q)$ are given, then for every choice of the numbers
$\lambda_{1},\lambda_{2},...$ from the gaps $\Delta_{1},\Delta_{2},...$ of the
spectrum of the Hill operator and for every choice of the signs of the numbers
$u_{-}(\lambda_{1}),$ $u_{-}(\lambda_{2}),...,$ there exist a potential $q$
having $\Lambda_{1}(t),$ $\Lambda_{2}(t),...$ as a band functions and
$\lambda_{1},\lambda_{2},...$ as the Dirichlet eigenvalues. In spite of this,
it is possible to determine uniquely (modulo (2)) the potential $q$ of the
multidimensional Schr\"{o}dinger operator $L(q)$ from only the given band
functions for a certain class of potential. Because, in the case $d>1$ the
band functions give more informations. Namely, the band functions give the
spectral invariants that have no meaning in the case $d=1$. We solve the
inverse problem by these spectral invariants. We will discuss this in the end
of the introduction.

The inverse problem for the multidimensional Schr\"{o}dinger operator $L(q)$
for the first time is investigated by Eskin, G., Ralston, J., Trubowitz, E. in
the papers [2,3]. In [2] it is proved the following result:

\textit{Assume that the lattice }$\Omega$ \textit{of }$\mathbb{R}^{d}$\textit{
is such that, for }$\omega,$ $\omega^{^{\prime}}\in\Omega,$\textit{ }%
$\mid\omega^{^{\prime}}\mid=\mid\omega\mid$\textit{\ }\ \textit{implies
}$\omega^{^{\prime}}=\pm\omega$\textit{. If }$q(x)$ \textit{and}
$\widetilde{q}(x)$\textit{ are real analytic, then the equality }%
\begin{equation}
Spec(L_{0}(q))=Spec(L_{0}(\widetilde{q}))
\end{equation}
\textit{implies the equalities }%
\begin{equation}
Spec(L_{t}(q))=Spec(L_{t}(\widetilde{q}))
\end{equation}
\textit{for all }$t\in\mathbb{R}^{d}$\textit{, where} $Spec(L_{t}(q))$
\textit{is the spectrum of the operator }$L_{t}(q)$\textit{ and }$L_{0}%
(q)$\textit{ is the operator }$L_{t}(q)$\textit{ when }$t=(0,0,...,0).$

In [3] it is proved the following result for the two-dimensional
Schr\"{o}dinger operator $L(q)$:

\textit{For }$\Omega\subset\mathbb{R}^{2}$\textit{\ satisfying the condition:
if }$\mid\omega^{^{\prime}}\mid=\mid\omega\mid$\textit{\ for }$\omega,$
$\omega^{^{\prime}}\in\Omega,$ \textit{then }$\omega^{^{\prime}}=\pm\omega
$\textit{; there is a set }$\{M_{\alpha}\}$\textit{\ of manifolds of
potentials such that }

\textit{a) }$\{M_{\alpha}:\alpha\in\lbrack0,1]\}$\textit{\ is dense in the set
of smooth periodic potentials in the }$C^{\infty}$-\textit{topology, }

\textit{b) for each }$\alpha$\textit{\ there is a dense open set }$Q_{\alpha
}\subset M_{\alpha}$\textit{\ such that for }$q\in Q_{\alpha}$ \textit{the set
of real analytic }$\widetilde{q}$\textit{\ satisfying (7)and the set of
}$\widetilde{q}\in C^{6}(F)$\textit{\ satisfying (8) for all} $t\in
\mathbb{R}^{2}$ \textit{are finite modulo translations in (2)}.

Eskin, G. [4] extend the results of the papers [2,3] to the case of
\ two-dimensional Schr\"{o}dinger operator
\[
H=(i\nabla+A(x))^{2}+V(x),\text{ }x\in\mathbb{R}^{2}%
\]
with periodic magnetic potential $A(x)=(A_{1}(x),A_{2}(x))$ and electric
potential $V(x).$ The proof of the results of the papers [2-4] is not
constructive and does not seem to give any idea about possibility to construct
explicitly a periodic potential.

In the paper [11] we constructed a set $D$\ of trigonometric polynomials of
the form
\begin{equation}
q(x)=%
%TCIMACRO{\tsum \limits_{a\in Q(N,M,S)}}%
%BeginExpansion
{\textstyle\sum\limits_{a\in Q(N,M,S)}}
%EndExpansion
z(a)e^{i\langle a,x\rangle},\ \
\end{equation}
where $Q(N,M,S)=:\{(n,m,s):|n|\leq N,\ |m|\leq M,\ |s|\leq S\}\backslash
\{(0,0,0)\}$ is a subset of the lattice $\mathbb{Z}^{3},$ satisfying the
following conditions:

$z(n,m,s)\neq0$ for $(n,m,s)\in B(N,M,S)\cup C(\sqrt{N})$ and

$z(n,m,s)=0$ for $\ (n,m,s)\in(Q(N,M,S))\backslash(C(\sqrt{N})\cup B(N,M,S)),$
where%
\begin{align*}
B(N,M,S)  &  =\{(n,m,s)\in Q(N,M,S):nms(|n|-N)(|m|-M)(|s|-S)=0\},\\
C(\sqrt{N})  &  =\left\{  (n,m,s):0<\mid n\mid<\tfrac{1}{2}\sqrt{N},0<\mid
m\mid<\tfrac{1}{2}\sqrt{N},0<\mid s\mid<\tfrac{1}{2}\sqrt{N}\right\}
\end{align*}
and $N$, $M$, $S$ are large prime numbers satisfying $S>2M,\ \ M>2N,\ \ N\gg
1$. Then we proved that: $D$ is dense in $W_{2}^{s}(\mathbb{R}^{3}/\Omega),$
where $s>3,$ in the $C^{\infty}$- topology and any element $q$ of the set $D$
can be determined constructively and uniquely, modulo inversion and
translations (2), from the given Bloch eigenvalues.

Thus, in the papers [2-4] and in our papers (in [11] and in this paper) the
different aspects of the inverse problem are investigated by absolutely
different methods. It follows from (3) and from the conditions on (9) that the
intersection of the set of potentials investigated in [11] and in this paper
is empty set. In this paper and in [11] we determine constructively the
potential $q$ of the three-dimensional Schr\"{o}dinger operator $L(q)$ from
the spectral invariants that were determined constructively in [10] from the
given band functions. As a result, we determine constructively the potential
from the given band functions. Actually, we do not only show how to construct
a periodic potential $q$ with the desired properties but even present an
algorithm for the construction. Moreover, in this paper we investigate the
stability of the algorithm and prove some uniqueness theorems which were not
done in [11].

To describe the brief scheme of this paper, we begin by recalling the
definition of some well-known concepts and the invariants obtained in [10]
that will be used here. An element $a$ of the lattice $\Gamma$ is said to be a
visible element of $\Gamma$ if $a$ is an element of $\Gamma$ of the minimal
norm belonging to the line $a\mathbb{R}.$ Denote by $S$ the set of all visible
elements of $\Gamma.$ Clearly,
\begin{equation}
q(x)=\frac{1}{2}%
%TCIMACRO{\tsum \limits_{a\in S}}%
%BeginExpansion
{\textstyle\sum\limits_{a\in S}}
%EndExpansion
q^{a}(x),
\end{equation}
where
\begin{equation}
q^{a}(x)=%
%TCIMACRO{\tsum \limits_{n\in\mathbb{Z}}}%
%BeginExpansion
{\textstyle\sum\limits_{n\in\mathbb{Z}}}
%EndExpansion
z(na)e^{in\langle a,x\rangle}.
\end{equation}
The function $q^{a}(x)$ defined in (11) are known as directional potential of
(10) corresponding to the visible element $a$. In Proposition 1 of Section 2
we prove that every element $a$ of $Q(1,1,1)$ is a visible element of $\Gamma$
and the directional potential (11) of (3) has the form
\begin{equation}
q^{a}(x)=z(a)e^{i\langle a,x\rangle}+z(-a)e^{-i\langle a,x\rangle}.
\end{equation}
Let $a$ be a visible element of $\Gamma,$ $\Omega_{a}$ be the sublattice
$\{\omega\in\Omega:\langle\omega,a\rangle=0\}$ of $\Omega$ in the hyperplane
$H_{a}=\{x\in\mathbb{R}^{3}:\langle x,a\rangle=0\}$ and%
\begin{equation}
\Gamma_{a}=:\{\gamma\in H_{a}:\langle\gamma,\omega\rangle\in2\pi
\mathbb{Z}\text{, }\forall\omega\in\Omega_{a}\}
\end{equation}
be the lattice dual to $\Omega_{a}.$ Let $\beta$ be a visible element of
$\Gamma_{a}$ and $P(a,\beta)$ be the plane containing $a$, $\beta$ and the
origin. Define a function $q_{a,\beta}(x)$ by
\begin{equation}
q_{a,\beta}(x)=\sum_{c\in(P(a,\beta)\cap\Gamma)\backslash a\mathbb{R}}\frac
{c}{\langle\beta,c\rangle}z(c)e^{i\langle c,x\rangle}.
\end{equation}
In the paper [10] we constructively determined the following spectral
invariants
\begin{equation}
I(a)=\ \int_{F}\left\vert q^{a}(x)\right\vert ^{2}dx,\
\end{equation}%
\begin{equation}
I_{1}(a,\beta)=\int_{F}\left\vert q_{a,\beta}(x)\right\vert ^{2}q^{a}(x)dx
\end{equation}
from the asymptotic formulas for the band functions of $L(q)$ obtained in
[8,9], where $q^{a}(x)$ is the directional potential (11). Moreover, in [10]
we constructively determined the invariant
\begin{equation}
I_{2}(a,\beta)=\int_{F}|q_{a,\beta}(x)|^{2}(z^{2}(a)e^{i2\langle a,x\rangle
}+z^{2}(-a)e^{-i2\langle a,x\rangle})dx
\end{equation}
when $q^{a}(x)$ has the form (12). Since all the directional potential of
\ (3) have the form (12) (see Proposition 1), we have the invariants (15)-(17)
for all $a\in Q(1,1,1).$

In Section 2 we describe the invariants (15)-(17) for (3).

In Section 3 fixing the inversion and translations (2), we give an algorithm
for the unique determination of the potential $q$ of the three-dimensional
Schr\"{o}dinger operator $L(q)$ from the invariants (15)-(17). Since the
invariants (16) and (17) do not exist in the case $d=1,$ we do not use the
investigations of the inverse problem for the one dimensional Schr\"{o}dinger
operator $H(q)$. For this reason, we do not discuss a great number of papers
about the inverse problem of the Hill operator.

In Section 4 we study the stability of the algorithm with respect to errors
both in the invariants (15)-(17) and in the Bloch eigenvalues. Note that we
determine constructively the potential from the band functions in two steps.
At the first step we determined the invariants from the band functions in the
paper [10]. At the second step, which is given in Section 3, we find the
potential from the invariants. In Section 4 we consider the stability of the
problems studied in the both steps. First, using the asymptotic formulas
obtained in [10], we write down explicitly the asymptotic expression of the
invariants (15)-(17) in terms of the band functions and consider the stability
of the invariants with respect to the errors in the Bloch eigenvalues (Theorem
5). Then we prove the stability of the algorithm given in Section 3 with
respect to the errors in the invariants (Theorem 6).

In Section 5 we prove some uniqueness theorems. First, we prove a theorem
about Hill operator $H(p)$ when $p(x)$ is a trigonometric polynomial (see
Theorem 7). Then we construct a set $W$ of all periodic functions $q(x)$ whose
directional potentials (see (10), (11)) $q^{a}(x)$ for all $a\in
S\backslash\{\gamma_{1},\gamma_{2},\gamma_{3}\}$ are arbitrary \ continuously
differentiable functions, where $S$ is the set of all visible elements of
$\Gamma$, $\{\gamma_{1},\gamma_{2},\gamma_{3}\}$ is a basis of $\Gamma$
satisfying (6), and the directional potentials $q^{\gamma_{1}}(x),$
$q^{\gamma_{2}}(x)$, $q^{\gamma3}(x)$ satisfy some conditions. At the end we
prove that if $q$ is of the form (3), $\widetilde{q}\in W$ and the band
functions of $L(q)$ and $L(\widetilde{q})$ coincide, then $\widetilde{q}$ is
equal to $q$ modulo inversion and translations (2) (see Theorem 8).

\section{On the spectral invariants (15)-(17)}

To describe the invariant (15) let us prove the following proposition.

\begin{proposition}
Every element $a$ of the set $Q(1,1,1),$ defined in (5), is a visible element
of $\Gamma$ and the corresponding directional potential (11) has the form (12).
\end{proposition}

\begin{proof}
Let $a$ be element of $Q(1,1,1).$ By the definition of $Q(1,1,1)$%
\begin{equation}
a=n\gamma_{1}+m\gamma_{2}+s\gamma_{3},\text{ }|n|\leq1,\ |m|\leq
1,\ |s|\leq1,\text{ }a\neq0.
\end{equation}
If $a$ is not a visible element of $\Gamma,$ then there exists a visible
element $b$ of $\Gamma$ such that $a=kb$ for some integer $k>1.$ This with
(18) implies that
\begin{equation}
b=\tfrac{1}{k}(n\gamma_{1}+m\gamma_{2}+s\gamma_{3}).
\end{equation}
Since $b\in\Gamma$ and $\{\gamma_{1},\gamma_{2},\gamma_{3}\}$ is a basis of
$\Gamma$ we have $b=n_{1}\gamma_{1}+m_{1}\gamma_{2}+s_{1}\gamma_{3},$ where
$n_{1},m_{1},s_{1}$ are integers. Combining this with (19) and taking into
account the linearly independence of the vectors $\gamma_{1},\gamma_{2}%
,\gamma_{3}$, we get
\[
(n_{1}-\tfrac{n}{k})\gamma_{1}+(m_{1}-\tfrac{m}{k})\gamma_{2}+(s_{1}-\tfrac
{s}{k})\gamma_{3}=0,\text{ and }n_{1}-\tfrac{n}{k}=\text{ }m_{1}-\tfrac{m}%
{k}=\text{ }s_{1}-\tfrac{s}{k}=0.
\]
This is impossible, since $|n|\leq1,\ |m|\leq1,\ |s|\leq1$, at least one of
the numbers $n,$ $m,\ s$ is not zero (see (18)), $k>1$ and the numbers
$n_{1},m_{1},s_{1}$ are integers. This contradiction shows that any element
$a$ of $Q(1,1,1)$ is a visible element of $\Gamma.$ Therefore, it follows from
the definition of $Q(1,1,1)$ (see (5)) that the line $a\mathbb{R}$ contains
only two elements $a$ and $-a$ of the set $Q(1,1,1).$ This means that the
directional potential (11) has the form (12)
\end{proof}

By Proposition 1 the invariant (15) for the potential (3) has the form
\begin{equation}
I(a)=|z(a)|^{2},\text{ }\forall a\in Q(1,1,1),
\end{equation}
that is, we determine the absolute value of $z(a)$ for all $a\in Q(1,1,1).$

To investigate the invariants (16) and (17), we use the conditions in (6).
Therefore, first, let us consider these conditions.

\begin{proposition}
Any lattice $\Gamma$ has a basis $\{\gamma_{1},\gamma_{2},\gamma_{3}\}$
satisfying (6). In particular, if%
\begin{equation}
\Gamma=\{(na,mb,sc):n,m,s\in\mathbb{Z}\},\text{ }%
\end{equation}
where $a,b,c\in\mathbb{R}\backslash\{0\},$ then at least one of the bases
$\{(a,0,0),(a,b,0),(a,b,c)\}$ and

$\{(-a,0,0),(a,b,0),(a,b,c)\}$ of $\Gamma$ satisfies (6).
\end{proposition}

\begin{proof}
Suppose that a basis $\{\gamma_{1},\gamma_{2},\gamma_{3}\}$ of $\Gamma$ does
not satisfy (6). Define $\{\widetilde{\gamma}_{1},\widetilde{\gamma}%
_{2},\widetilde{\gamma}_{3}\}$ by%
\[
\widetilde{\gamma}_{1}=\gamma_{1},\text{ }\widetilde{\gamma}_{2}=n\gamma
_{1}+\gamma_{2},\text{ }\widetilde{\gamma}_{3}=m\gamma_{1}+s\gamma_{2}%
+\gamma_{3},
\]
where $n,m,s$ are integers. Since $\gamma_{1}=\widetilde{\gamma}_{1},$
$\gamma_{2}=\widetilde{\gamma}_{2}-n\widetilde{\gamma}_{1},$ $\gamma
_{3}=\widetilde{\gamma}_{3}-m\widetilde{\gamma}_{1}-s(\widetilde{\gamma}%
_{2}-n\widetilde{\gamma}_{1}),$ the triple $\{\widetilde{\gamma}%
_{1},\widetilde{\gamma}_{2},\widetilde{\gamma}_{3}\}$ is also basis of
$\Gamma.$ In (6) replacing $\{\gamma_{1},\gamma_{2},\gamma_{3}\}$ by
$\{\widetilde{\gamma}_{1},\widetilde{\gamma}_{2},\widetilde{\gamma}_{3}\}$, we
obtain\ $12$ inequalities with respect to $n,m$ and $s$. Since $n,m$ and $s$
are arbitrary integers one can readily see that there exists $n,m$ and $s$ for
which these inequalities hold. For example, let
\begin{equation}
\widetilde{\gamma}_{1}=\gamma_{1},\text{ }\widetilde{\gamma}_{2}=n\gamma
_{1}+\gamma_{2},\text{ }\widetilde{\gamma}_{3}=n^{2}\gamma_{1}+\gamma_{3},
\end{equation}
where $n$ is a large positive number, that is, $n\gg1$. Then it follows from
(22) that
\[
\left\langle \widetilde{\gamma}_{i},\text{ }\widetilde{\gamma}_{j}%
\right\rangle \gg1,\text{ }\left\langle \widetilde{\gamma}_{i}+\widetilde
{\gamma}_{j},\widetilde{\gamma}_{j}\right\rangle \gg1,\text{ }\forall i\neq
j,
\]
that is, the first and second inequalities in (6) hold. Besides, by (22), we
have
\begin{equation}
\mid\widetilde{\gamma}_{1}\mid^{2}\sim1,\text{ }\mid\widetilde{\gamma}_{2}%
\mid^{2}\sim n^{2},\text{ }\mid\widetilde{\gamma}_{3}\mid^{2}\sim n^{4},\text{
}\left\langle \widetilde{\gamma}_{i},\text{ }\widetilde{\gamma}_{j}%
\right\rangle =O(n^{3}),
\end{equation}
where \ $a_{n}\sim b_{n}$ means that there exist positive constants $c_{1}$
and $c_{2}$ such that

$c_{1}\mid b_{n}\mid<\mid a_{n}\mid<c_{2}\mid b_{n}\mid$, for $\ n=1,2,...$.
The third inequality of (6) holds due to (23). By (23) the term $\pm
\mid\widetilde{\gamma}_{3}\mid^{2}$ in the fourth inequality of (6) can not be
canceled by the other terms of this inequality. Thus, we proved that any
lattice $\Gamma$ has a basis $\{\widetilde{\gamma}_{1},\widetilde{\gamma}%
_{2},\widetilde{\gamma}_{3}\}$ satisfying (6).

Note that, for the given lattice, one can easily find the basis satisfying
(6). For example, in the case (21), one can readily see that the basis
$\{(a,0,0),(a,b,0),(a,b,c)\}$ satisfies (6) if $c^{2}\neq3a^{2}$ and the basis
$\{(-a,0,0),(a,b,0),(a,b,c)\}$ satisfies (6) if $c^{2}\neq a^{2}.$ Thus at
least one of the bases $\{(a,0,0),(a,b,0),(a,b,c)\}$ and
$\{(-a,0,0),(a,b,0),(a,b,c)\}$ satisfies (6)
\end{proof}

Now to describe the invariants (16) and (17) for (3) let us introduce some
notations. If $b\in(\Gamma\cap P(a,\beta))\backslash a\mathbb{R}$, then the
plane $P(a,\beta)$ coincides with the plane $P(a,b)$. Moreover, every vector
$b\in(P(a,\beta)\cap\Gamma)\backslash a\mathbb{R}$ has an orthogonal
decomposition (see (20) in [8])%
\begin{equation}
b=s\beta+\mu a,
\end{equation}
where $s$ is a nonzero integer, $\beta$ is a visible element of $\Gamma_{a}$
(see (13)) and $\mu$ is a real number. Therefore, for every plane $P(a,b)$,
where $b\in\Gamma,$ there exists a plane $P(a,\beta)$, where $\beta$ is
defined by (24), coinciding with $P(a,b).$ For every pair $\{a,b\}$, where $a$
is visible element of $\Gamma$ and $b\in\Gamma,$ we redenote by $I_{1}(a,b)$
and $I_{2}(a,b)$ the invariants $I_{1}(a,\beta)$ and $I_{2}(a,\beta)$ defined
in (16) and (17) respectively, where $\beta$ is a visible element of
$\Gamma_{a}$ defined by (24).

\begin{theorem}
The following equalities for the invariant (16) hold:
\begin{equation}
I_{1}(\gamma_{i}+\gamma_{j},\gamma_{i})=A_{1}(\gamma_{i}+\gamma_{j},\gamma
_{i})\operatorname{Re}(z(-\gamma_{i}-\gamma_{j})z(\gamma_{j})z(\gamma_{i})),
\end{equation}%
\begin{equation}
I_{1}(\gamma_{i}-\gamma_{j},\gamma_{i})=A_{1}(\gamma_{i}-\gamma_{j},\gamma
_{i})\operatorname{Re}(z(-\gamma_{i}+\gamma_{j})z(-\gamma_{j})z(\gamma_{i})),
\end{equation}%
\begin{equation}
I_{1}(\gamma,\gamma_{i})=A_{1}(\gamma,\gamma_{i})\operatorname{Re}%
(z(-\gamma)z(\gamma-\gamma_{i})z(\gamma_{i})),
\end{equation}%
\begin{equation}
I_{1}(2\gamma_{i}-\gamma,\gamma_{i})=A_{1}(2\gamma_{i}-\gamma,\gamma
_{i})\operatorname{Re}(z(\gamma-2\gamma_{i})z(\gamma_{i}-\gamma)z(\gamma
_{i})),
\end{equation}
where $A_{1}(\gamma_{i}\pm\gamma_{j},\gamma_{i}),$ $A_{1}(\gamma,\gamma_{i}%
),$\ $A_{1}(2\gamma_{i}-\gamma,\gamma_{i})$ are nonzero numbers defined by
\begin{equation}
A_{1}(a,b)=2\left(  (\left\langle b,\beta\right\rangle )^{-2}+(\left\langle
a-b,\beta\right\rangle )^{-2}\right)  \left\langle a-b,b\right\rangle ,
\end{equation}
$\{\gamma_{1},\gamma_{2},\gamma_{3}\}$ is a basis of $\Gamma$ satisfying (6),
$\gamma=\gamma_{1}+\gamma_{2}+\gamma_{3}$ and $\operatorname{Re}(z)$ is the
real part of $z.$
\end{theorem}

\begin{proof}
If the potential $q(x)$ has the form (3), then (14) becomes
\begin{equation}
q_{a,\beta}(x)=\sum_{c\in(P(a,\beta)\cap Q)\backslash a\mathbb{R}}\frac
{c}{\langle\beta,c\rangle}z(c)e^{i\langle c,x\rangle},
\end{equation}
where, for brevity, $Q(1,1,1)$ is denoted by $Q.$ Using this and (11) in (16)
and taking into account that the invariant $I_{1}(a,\beta)$ defined by (16) is
redenoted by $I_{1}(a,b),$ we get
\begin{equation}
I_{1}(a,b)=\Sigma_{1}+\Sigma_{2},
\end{equation}
where
\[
\Sigma_{1}=\sum\limits_{c\in(P(a,b)\cap Q)\backslash a\mathbb{R}}%
\frac{\left\langle c,c+a\right\rangle }{\left\langle c,\beta\right\rangle
\left\langle c+a,\beta\right\rangle }z(c)z(-a-c)z(a),
\]%
\[
\Sigma_{2}=\sum\limits_{c\in(P(a,b)\cap Q)\backslash a\mathbb{R}}%
\frac{\left\langle c,c-a\right\rangle }{\left\langle c,\beta\right\rangle
\left\langle c-a,\beta\right\rangle }z(c)z(a-c)z(-a)
\]
and $\beta$ is a visible element of $\Gamma_{a}$ defined by (24). Since
$Q(1,1,1)$ is symmetric with respect to the origin, the substitution
$\widetilde{c}=-c$ in $\Sigma_{1}$ does not change $\Sigma_{1}$. Using this
substitution in $\Sigma_{1}$ and then taking into account that
$z(-b)=\overline{z(b)},$ $\left\langle a,\beta\right\rangle =0,$ we obtain
$\Sigma_{1}=\overline{\Sigma_{2}}$. This with (31) gives
\begin{equation}
I_{1}(a,b)=2\operatorname{Re}\left(  z(-a)\left(  \sum\limits_{c\in(P(a,b)\cap
Q)\backslash a\mathbb{R}}\frac{\left\langle a-c,c\right\rangle }{(\left\langle
c,\beta\right\rangle )^{2}}z(a-c)z(c)\right)  \right)  .
\end{equation}
Since $a$, $\beta$, $(0,0,0)$ belong to the plane $P(a,b)$ and $\beta$
orthogonal to the line $a\mathbb{R},$ we have
\begin{equation}
\left\langle c,\beta\right\rangle \neq0,\text{ }\forall c\in(P(a,b)\cap
Q)\backslash a\mathbb{R}.
\end{equation}

Now using \ (32) we obtain the invariants (25) and (26) as follows. First let
us consider (25). Let $a=\gamma_{i}+\gamma_{j}$ and $b=\gamma_{i}.$ Then
\[
(P(a,b)\cap Q)\backslash a\mathbb{R}=\{\pm\gamma_{i},\text{ }\pm\gamma
_{j},\text{ }\pm(\gamma_{i}-\gamma_{j})\}.
\]
One the other hand, if $c\in\{-\gamma_{i},$ $-\gamma_{j},$ $\pm(\gamma
_{i}-\gamma_{j})\},$ then $a-c\notin Q.$ Therefore, the summation in the
formula (32) for the case $a=\gamma_{i}+\gamma_{j},$ $b=\gamma_{i}$ is taken
over $c\in\{\gamma_{i},$ $\gamma_{j}\}$ and hence (25) holds. It follows from
(33) and from the first inequality in (6) that $A_{1}(\gamma_{i}+\gamma
_{j},\gamma_{i})\neq0.$

Replacing $a=\gamma_{j}$ by $-\gamma_{j}$ and arguing as in the proof of (25),
we get (26).

Now let us consider (27). Let $a=\gamma=\gamma_{1}+\gamma_{2}+\gamma_{3}$ and
$b=\gamma_{1}.$ Then
\[
(P(a,b)\cap Q)\backslash a\mathbb{R}=\{\pm\gamma_{1},\text{ }\pm(\gamma
_{2}+\gamma_{3})\}.
\]
One the other hand, if $c=-\gamma_{1},$ or $c=-\gamma_{2}-\gamma_{3},$ then
$a-c\notin Q.$ Therefore, the summation in the formula (32) for this case is
taken over $c\in\{\gamma_{1},$ $\gamma_{2}+\gamma_{3}\}$ and hence (27) holds
for $i=1$. In the same way, we obtain (27) for $i=2,3.$

Now let us consider (28). Let $a=2\gamma_{i}-\gamma$ and $b=\gamma_{i}.$ Then
\[
(P(a,b)\cap Q)\backslash a\mathbb{R}=\{\pm\gamma_{i},\text{ }\pm(\gamma
_{i}-\gamma)\}.
\]
One the other hand, if $c=-\gamma_{i},$ or $c=\gamma-\gamma_{i},$ then
$a-c\notin Q.$ Therefore, the summation in the formula (32) for this case is
taken over $c\in\{\gamma_{i},\gamma_{i}-\gamma\}$ and hence (28) holds. Since
$\gamma_{i}-\gamma=-(\gamma_{j}+\gamma_{k}),$ it follows from the second
inequality in (6) that $A_{1}(2\gamma_{i}-\gamma,\gamma_{i})\neq0$.
\end{proof}

\begin{theorem}
The following equalities for the invariant (17) hold:
\begin{equation}
I_{2}(\gamma_{i},\gamma_{j})=A_{2}(\gamma_{i},\gamma_{j})\operatorname{Re}%
(z^{2}(-\gamma_{i})z(\gamma_{i}+\gamma_{j})z(\gamma_{i}-\gamma_{j})),
\end{equation}%
\begin{equation}
I_{2}(\gamma_{i},\gamma-\gamma_{i})=A_{2}(\gamma_{i},\gamma-\gamma
_{i})\operatorname{Re}(z^{2}(-\gamma_{i})z(\gamma)z(2\gamma_{i}-\gamma)),
\end{equation}
where $A_{2}(\gamma_{i},\gamma_{j})$, $A_{2}(\gamma_{i},\gamma-\gamma_{i})$
are nonzero numbers defined by

$A_{2}(a,b)=2(a-b,a+b)(b,\beta)^{-2}$ and $\gamma,$ $\gamma_{1},$ $\gamma
_{2},$ $\gamma_{3}$ are defined in Theorem 1.
\end{theorem}

\begin{proof}
Replacing $a$ by $2a,$ and arguing as in the proof of (32), we get
\begin{equation}
I_{2}(a,b)=2\operatorname{Re}\left(  z^{2}(-a)\left(
%TCIMACRO{\tsum \limits_{c\in(P(a,b)\cap Q)\backslash a\mathbb{R}}}%
%BeginExpansion
{\textstyle\sum\limits_{c\in(P(a,b)\cap Q)\backslash a\mathbb{R}}}
%EndExpansion
\frac{\left\langle 2a-c,c\right\rangle }{(\left\langle c,\beta\right\rangle
)^{2}}z(2a-c)z(c)\right)  \right)  .
\end{equation}
In (36) replacing $c$ by $a+c$ and taking into account that $\left\langle
a,\beta\right\rangle =0$,\ we obtain the invariant%
\begin{equation}
I_{2}(a,b)=2\operatorname{Re}\left(  z^{2}(-a)\left(  \sum\limits_{c\in
(P(a,b)\cap Q)\backslash a\mathbb{R}}\frac{\left\langle a+c,a-c\right\rangle
}{(\left\langle c,\beta\right\rangle )^{2}}z(a+c)z(a-c)\right)  \right)  .
\end{equation}

Now using this, we obtain the invariants (34) and (35) as follows. First let
us consider (34). Let $a=$ $\gamma_{i},\ b=\gamma_{j}.$ Then
\[
(P(a,b)\cap Q)\backslash a\mathbb{R}=\{\pm\gamma_{j},\text{ }\pm(\gamma
_{i}-\gamma_{j}),\text{ }\pm(\gamma_{i}+\gamma_{j})\}.
\]
One the other hand, if $c=\pm(\gamma_{i}-\gamma_{j}),$ or $c=\pm(\gamma
_{i}+\gamma_{j}),$ then at least one of the vectors $a-c$ and $a+c$ does not
belong to $Q.$ Therefore, the summation in (37) for this case is taken over
$c\in\{\pm\gamma_{j}\}$ and hence (34) holds. By the third inequality in (6)
we have $A_{2}(\gamma_{i},\gamma_{j})\neq0.$

Now let us consider (35). Let $a=\gamma_{i}$ and $b=\gamma-\gamma_{i}.$ Then
\[
(P(a,b)\cap Q)\backslash a\mathbb{R}=\{\pm\gamma,\text{ }\pm(\gamma-\gamma
_{i}),\text{ }\pm(\gamma-2\gamma_{i})\}.
\]
If $c=\gamma,$ then $c+a=\gamma+\gamma_{i}\notin Q.$ If $c=-\gamma,$ then
$c-a=-\gamma-\gamma_{i}\notin Q.$ If $c=\gamma-2\gamma_{i},$ then
$c-a=\gamma-3\gamma_{i}\notin Q.$ If $c=-(\gamma-2\gamma_{i}),$ then
$c+a=-\gamma+3\gamma_{i}\notin Q.$ Therefore, the summation in the formula
(37) for this case is taken over $c\in\{\pm(\gamma-\gamma_{i})\}$ and hence
(35) holds. Since $\gamma=\gamma_{i}+\gamma_{j}+\gamma_{k},$ it follows from
the last inequality in (6) that $A_{2}(\gamma_{i},\gamma-\gamma_{i})\neq0.$
\end{proof}

\section{Finding the potential from the invariants}

In this section we give an algorithm for finding the all Fourier coefficients
$z(a)$ of the potential (3) from the invariants (25)-(28), (34) and (35).
First, let us introduce some notations. The number of elements of the set
\[
\{n\gamma_{1}+m\gamma_{2}+s\gamma_{3}:|n|\leq1,\ |m|\leq1,\ |s|\leq1\}
\]
is $27,$ since the numbers $n,m,s$ take $3$ values $-1,0,1$ independently. The
set $Q(1,1,1)$ (see (5)) is obtained from this set by eliminating the element
$(0,0,0),$ and hence consist of $26$ elements. Moreover, if $\gamma\in
Q(1,1,1),$ then $-\gamma\in Q(1,1,1)$ and $\gamma\neq-\gamma.$ Hence the
elements of $Q(1,1,1)$ can be denoted by $\gamma_{1},\gamma_{2},...,\gamma
_{13}$ and $-\gamma_{1},-\gamma_{2},...,-\gamma_{13}.$ Let us denote the
elements $\gamma_{1},\gamma_{2},...,\gamma_{7}$ as following: $\gamma
_{1},\gamma_{2},\gamma_{3}$ be a basis of $\Gamma$ satisfying (6) and
\begin{equation}
\gamma_{4}=\gamma_{2}+\gamma_{3},\text{ }\gamma_{5}=\gamma_{1}+\gamma
_{3},\text{ }\gamma_{6}=\gamma_{1}+\gamma_{2},\text{ }\gamma_{7}=\gamma
_{1}+\gamma_{2}+\gamma_{3}.
\end{equation}
Introduce the notations
\begin{equation}
z(\gamma_{j})=a_{j}+ib_{j}=r_{j}e^{i\alpha_{j}},
\end{equation}
where $a_{j}\in\mathbb{R}$, $b_{j}\in\mathbb{R}$, $r_{j}=\mid z(\gamma
_{j})\mid\in(0,\infty),$ and $\alpha_{j}=\alpha(\gamma_{j})=\arg(z(\gamma
_{j}))\in\lbrack0,2\pi)$ for

$i=1,2,...,13$. Since the modulus $r_{j}$ of the Fourier coefficients
$z(\gamma_{j})$ are known due to (20), we need to know the values of the
arguments $\alpha_{j}$ of $z(\gamma_{j}).$ For this we use the following
conditions on the arguments $\alpha_{1},\alpha_{2},..,\alpha_{7}:$
\begin{align}
\alpha_{7}-\alpha_{1}-\alpha_{2}-\alpha_{3}  &  \neq\pi k,\text{ }\alpha
_{7}-\alpha_{s+3}-\alpha_{s}\neq\pi k,\text{ }\alpha_{m+3}-\alpha_{j+3}%
+\alpha_{m}-\alpha_{j}\neq\pi k,\nonumber\\
\alpha_{4}-\alpha_{2}-\alpha_{3}  &  \neq\frac{\pi}{2}k,\text{ }\alpha
_{5}-\alpha_{1}-\alpha_{3}\neq\frac{\pi}{2}k,\text{ }\alpha_{6}-\alpha
_{1}-\alpha_{2}\neq\frac{\pi}{2}k,\\
\alpha_{4}+\alpha_{5}-\alpha_{1}-\alpha_{2}-2\alpha_{3}  &  \neq\pi k,\text{
}\alpha_{4}+\alpha_{6}-\alpha_{1}-\alpha_{3}-2\alpha_{2}\neq\pi k,\text{
}\nonumber\\
\alpha_{5}+\alpha_{6}-\alpha_{2}-\alpha_{3}-2\alpha_{1}  &  \neq\pi
k,\nonumber
\end{align}
where $s=1,2,3;$ $k\in\mathbb{Z}$ and $m,j$ are integers satisfying $1\leq
m<j\leq3.$ In this section we give an algorithm for the unique (modulo (2))
determination of the potentials $q$ of the form (3) satisfying (40) from the
invariants (15)-(17). In the following remark we consider geometrically the
set of all potentials of the form (3) satisfying (40).

\begin{remark}
Since $z(\gamma)=\overline{z(-\gamma)}$, there exists one to one
correspondence between the trigonometric polynomials of the form (3) and the
vectors $(r_{1},\alpha_{1},r_{2},\alpha_{2},...,r_{13},\alpha_{13})$ of the
subset
\[
S=:(0,\infty)^{13}\otimes\lbrack0,2\pi)^{13}%
\]
of the space $\mathbb{R}^{26}.$ We use conditions (40) as restrictions on the
potential (3) and hence on the set $S.$ Denote by $S^{^{\prime}}$ the subset
of $S$ corresponding to the set of the potential (3) satisfying conditions
(40). The conditions (40) means that we eliminate from the subset
\[
D=:\{(\alpha_{1},\alpha_{2},...,\alpha_{7}):\alpha_{1}\in\lbrack0,2\pi
),\alpha_{1}\in\lbrack0,2\pi),\text{ }\alpha_{2}\in\lbrack0,2\pi),...,\text{
}\alpha_{7}\in\lbrack0,2\pi)\}
\]
of $\mathbb{R}^{7}$ the following six-dimensional hyperplanes
\begin{align*}
\{\alpha_{7}-\alpha_{1}-\alpha_{2}-\alpha_{3}  &  =\pi k\},\text{ }%
\{\alpha_{7}-\alpha_{s+3}-\alpha_{s}=\pi k\},\text{ }\{\alpha_{m+3}%
-\alpha_{j+3}+\alpha_{m}-\alpha_{j}=\pi k\},\\
\{\alpha_{4}-\alpha_{2}-\alpha_{3}  &  =\frac{\pi}{2}k\},\text{ }\{\alpha
_{5}-\alpha_{1}-\alpha_{3}=\frac{\pi}{2}k\},\text{ }\{\alpha_{6}-\alpha
_{1}-\alpha_{2}=\frac{\pi}{2}k\},\\
\{\alpha_{4}+\alpha_{5}-\alpha_{1}-\alpha_{2}-2\alpha_{3}  &  =\pi k\},\text{
}\{\alpha_{4}+\alpha_{6}-\alpha_{1}-\alpha_{3}-2\alpha_{2}=\pi k\},\\
\text{ }\{\alpha_{5}+\alpha_{6}-\alpha_{2}-\alpha_{3}-2\alpha_{1}  &  =\pi k\}
\end{align*}
of $\mathbb{R}^{7}=\{(\alpha_{1},\alpha_{2},...,\alpha_{7})\},$ where
$s=1,2,3;$ $k\in\mathbb{Z}$ and $m,j$ are integers satisfying $1\leq
m<j\leq3.$ In this notation we have
\[
S=(0,\infty)^{13}\otimes\lbrack0,2\pi)^{6}\otimes D,\text{ }S^{^{\prime}%
}=(0,\infty)^{13}\otimes\lbrack0,2\pi)^{6}\otimes D^{^{\prime}},
\]
where $D^{^{\prime}}$ is obtained from $D$ by eliminating the above
six-dimensional hyperplanes. It is clear that the $26$ \ dimensional measure
of the set $S\backslash S^{^{\prime}}$ is zero. Since the main result (Theorem
4) of this section is concerned to the potentials corresponding to the set
$S^{^{\prime}},$ we investigate the almost all potentials of the form (3).
\end{remark}

Since the operators $L(q(x-\tau))$ for $\tau\in F$ have the same band
functions, we may fix $\tau$, that is, take one of the functions $q(x-\tau),$
which determines three of the arguments. \ 

\begin{theorem}
There exists a unique value of $\tau\in F$ such that the following conditions
hold
\begin{equation}
\alpha(\tau,\gamma_{1})=\alpha(\tau,\gamma_{2})=\alpha(\tau,\gamma_{3})=0,
\end{equation}
where $\{\gamma_{1},\gamma_{2},\gamma_{3}\}$ is a basis of the lattice
$\Gamma$ and $\alpha(\tau,\gamma)=\arg(q(x-\tau),e^{i\langle\gamma,x\rangle
}).$
\end{theorem}

\begin{proof}
Let $\omega_{1},\omega_{2},\omega_{3}$ be a basis of $\Omega$ satisfying
\begin{equation}
\langle\gamma_{i},\omega_{j}\rangle=2\pi\delta_{i,j}%
\end{equation}
and $F=\{c_{1}\omega_{1}+c_{2}\omega_{2}+c_{3}\omega_{3}:c_{k}\in
\lbrack0,1),k=1,2,3\}$ be a fundamental domain $\mathbb{R}^{3}/\Omega$ of
$\Omega.$ If $\tau\in F,$ then we have\ $\tau=c_{1}\omega_{1}+c_{2}\omega
_{2}+c_{3}\omega_{3}.$ Therefore, using the notations of (3), (4) and (41) one
can readily see that
\begin{equation}
\alpha(\tau,\gamma)=\arg(q(x-\tau),e^{i\langle\gamma,x-\tau\rangle}%
e^{i\langle\gamma,\tau\rangle})=\alpha(\gamma)-\langle\gamma,\tau\rangle.
\end{equation}
This with (42) yields $\alpha(\tau,\gamma_{k})=\alpha(\gamma_{k})-2\pi c_{k}$
which means that (41) is equivalent to

$2\pi c_{k}=\alpha(\gamma_{k}),$ where $\alpha(\gamma_{k})\in\lbrack0,2\pi),$
$2\pi c_{k}\in\lbrack0,2\pi)$ and $k=1,2,3.$ Thus, there exists a unique value
of $\tau=c_{1}\omega_{1}+c_{2}\omega_{2}+c_{3}\omega_{3}\in F$ satisfying (41)
\end{proof}

By Theorem 3, without loss of generality, it can be assumed that
\begin{equation}
\alpha_{1}=\alpha_{2}=\alpha_{3}=0.
\end{equation}
Thus $z(\gamma_{i})=\mid z(\gamma_{i})\mid$ and by (20) $z(\gamma_{i})$ for
$i=1,2,3$ are the known positive numbers:%
\begin{equation}
z(\gamma_{i})=a_{i}>0,\text{ }\forall i=1,2,3.
\end{equation}
Using (43) one can easily verify that the expressions in the left-hand sides
of the inequalities in (40) do not depend on $\tau.$ Therefore, using the
assumption (44) one can readily see that the condition (40) has the form
\begin{equation}
\alpha_{7}\neq\pi k,\text{ }\alpha_{s}\neq\frac{\pi}{2}k,\text{ }\alpha
_{7}-\alpha_{s}\neq\pi k,\text{ }\alpha_{m}\pm\alpha_{j}\neq\pi k,
\end{equation}
where $k\in\mathbb{Z};$ $s=4,5,6;$ $j=4,5,6$; $m=4,5,6$ and $m\neq j.$ Using
the notation of (39) and taking into account that $r_{j}r_{m}\sin(\alpha
_{j}\pm\alpha_{m})=b_{j}a_{m}\pm b_{m}a_{j},$ $r_{j}r_{m}\neq0$ (see (4)), we
see that (46) can be written in the form
\begin{equation}
b_{7}\neq0,\text{ }a_{s}b_{s}\neq0,\text{ }b_{7}a_{s}-a_{7}b_{s}\neq0,\text{
}b_{j}a_{m}\pm b_{m}a_{j}\neq0, \tag{46.1}%
\end{equation}
where $s=4,5,6;$ $j=4,5,6$; $m=4,5,6$ and $m\neq j.$

The equality $(q(-x),e^{i\langle a,x\rangle})=\overline{(q(x),e^{i\langle
a,x\rangle})}$ shows that the imaginary part of the Fourier coefficients of
$q(x)$ and $q(-x)$ take the opposite values. Therefore, taking into account
the first inequality of (46.1), for fixing the inversion $q(x)\longrightarrow
q(-x),$ in the set of potentials of the form (3) satisfying (40), we assume
that%
\begin{equation}
b_{7}>0.
\end{equation}
Now using (44), (46.1), (47) and the invariants (25)-(28), (34), (35), we will
find the Fourier coefficients $z(a)$ for all $a\in Q.$

\begin{theorem}
The invariants (15)-(17) determine constructively and uniquely, modulo
inversion and translation (2), all the potentials of the form (3) satisfying (40).
\end{theorem}

\begin{proof}
To determine the potential (3), we find its Fourier coefficients step by step.

\textit{Step 1. }In this step using (25), (27), (46.1) and (47), we find%
\begin{equation}
z(\gamma_{1}+\gamma_{2}),\text{ }z(\gamma_{1}+\gamma_{3}),\text{ }z(\gamma
_{2}+\gamma_{3}),\text{ }z(\gamma_{1}+\gamma_{2}+\gamma_{3}).
\end{equation}
Since $z(\gamma_{1}),$ $z(\gamma_{2}),$ $z(\gamma_{3})$ are known positive
numbers (see (45)), the invariants in (25) give the real parts of the Fourier
coefficients $z(\gamma_{2}+\gamma_{3}),z(\gamma_{1}+\gamma_{3}),z(\gamma
_{1}+\gamma_{2}).$ Then, using (20), we find\ the absolute values of the
imaginary parts of these Fourier coefficients. Thus due to the notations of
(38) and (39), we have
\begin{equation}
z(\gamma_{2}+\gamma_{3})=a_{4}+it_{4}\mid b_{4}\mid,\text{ }z(\gamma
_{1}+\gamma_{3})=a_{5}+it_{5}\mid b_{5}\mid,\text{ }z(\gamma_{1}+\gamma
_{2})=a_{6}+it_{6}\mid b_{6}\mid,
\end{equation}
where $a_{m}$ and $\mid b_{m}\mid$ for $m=4,5,6$ are known real numbers and
$t_{m}$ is the sign of $b_{m},$ that is, is either $-1$ or $1$. To determine
$t_{4},t_{5},t_{6},$ we use (27). Using (45), (49) and the notations
$\gamma=\gamma_{1}+\gamma_{2}+\gamma_{3}=\gamma_{7},$ $z(\gamma_{7}%
)=a_{7}+ib_{7}$ (see (38), (39)) one sees that (27) for $i=1$ give us the
value of $\operatorname{Re}(a_{7}-ib_{7})(a_{4}+it_{4}\mid b_{4}\mid)a_{1}.$
In other word, we have the equation
\begin{equation}
a_{4}a_{7}+t_{4}\mid b_{4}\mid b_{7}=c_{1}%
\end{equation}
with respect to the unknowns $a_{7}$ and $b_{7},$ where $c_{1}$\ is the known
constant, since (27) is a given invariant. Here and in the forthcoming
equations by $c_{k}$ for $k=1,2,...$ we denote the known constants. In the
same way, from (27) for $i=2,3,$ we obtain
\begin{equation}
a_{5}a_{7}+t_{5}\mid b_{5}\mid b_{7}=c_{2},
\end{equation}%
\begin{equation}
a_{6}a_{7}+t_{6}\mid b_{6}\mid b_{7}=c_{3}.
\end{equation}
By (46.1) $t_{5}\mid b_{5}\mid a_{4}-t_{4}\mid b_{4}\mid a_{5}\neq0,$
$t_{6}\mid b_{6}\mid a_{4}-t_{4}\mid b_{4}\mid a_{6}\neq0,$ $t_{6}\mid
b_{6}\mid a_{5}-t_{5}\mid b_{5}\mid a_{6}\neq0.$ Therefore finding $b_{7}$
from the systems of equations generated by pairs \{(50), (51)\},

\{(50), (52)\}, \{(51), (52)\}, and taking into account (47), we get the
inequalities
\begin{equation}
\frac{a_{4}c_{2}-a_{5}c_{1}}{t_{5}\mid b_{5}\mid a_{4}-t_{4}\mid b_{4}\mid
a_{5}}>0,\text{ }\frac{a_{4}c_{3}-a_{6}c_{1}}{t_{6}\mid b_{6}\mid a_{4}%
-t_{4}\mid b_{4}\mid a_{6}}>0,\text{ }\frac{a_{5}c_{3}-a_{6}c_{2}}{t_{6}\mid
b_{6}\mid a_{5}-t_{5}\mid b_{5}\mid a_{6}}>0
\end{equation}
respectively. Now we prove that the relations (50)-(53) determines uniquely
the unknowns $a_{7}$,$b_{7},t_{4},t_{5},t_{6}.$ Suppose to the contrary that
there exists to different solutions $(a_{7},b_{7},t_{4},t_{5},t_{6})$ and
$(a_{7}^{^{\prime}}$,$b_{7}^{^{\prime}},t_{4}^{^{\prime}},t_{5}^{^{\prime}%
},t_{6}^{^{\prime}})$ of (50)-(53). Clearly, if $2$ components of the triple
$(t_{4}^{^{\prime}},t_{5}^{^{\prime}},t_{6}^{^{\prime}})$ take the opposite
values of the corresponding components of the triple $(t_{4},t_{5},t_{6})$
then all the inequalities in (53) do not hold simultaneously. Therefore, at
least, two component of $(t_{4}^{^{\prime}},t_{5}^{^{\prime}},t_{6}^{^{\prime
}})$ must be the same with the corresponding two components of $(t_{4}%
,t_{5},t_{6}).$ It can be assumed, without loss of generality, that
$t_{4}^{^{\prime}}=t_{4}$ and $t_{5}^{^{\prime}}=t_{5}.$ Then it follows from
the system of equation (50), (51) that $a_{7}^{^{\prime}}=a_{7}$,
$b_{7}^{^{\prime}}=b_{7}.$ Since $b_{6}b_{7}\neq0$ due to (46.1) it follows
from (52) that $t_{6}^{^{\prime}}=t_{6}.$ Thus the Fourier coefficients in
(48) can be determined from (50)-(53):
\begin{align}
z(\gamma_{2}+\gamma_{3})  &  =a_{4}+ib_{4},\text{ }z(\gamma_{1}+\gamma
_{3})=a_{5}+ib_{5},\text{ }z(\gamma_{1}+\gamma_{2})=a_{6}+ib_{6},\\
z(\gamma_{1}+\gamma_{2}+\gamma_{3})  &  =a_{7}+ib_{7}.
\end{align}

\textit{Step 2. }In this step using (34) and (46.1), we find%
\begin{equation}
z(\gamma_{1}-\gamma_{2}),\text{ }z(\gamma_{1}-\gamma_{3}),\text{ }z(\gamma
_{2}-\gamma_{3}).
\end{equation}
Writing (34) for $i=1,$ $j=2$ and taking into account that $z(-\gamma
_{i})=z(\gamma_{i})=a_{i}$ (see (45)) and $z(\gamma_{1}+\gamma_{2}%
)=a_{6}+ib_{6}$ (see (54)), we find the value of $a_{1}^{2}\operatorname{Re}%
(a_{6}+ib_{6})z(\gamma_{1}-\gamma_{2}).$ In other word, we have an equation
\begin{equation}
a_{6}x-b_{6}y=c_{4},
\end{equation}
where $z(\gamma_{1}-\gamma_{2})=x+iy.$ From (34) for $i=2,$ $j=1,$ in the same
way, we get%
\begin{equation}
a_{6}x+b_{6}y=c_{5}.
\end{equation}
Since $a_{6}b_{6}\neq0$ due to (46.1), from (57) and (58), we find $x$ and $y$
and hence $z(\gamma_{1}-\gamma_{2}).$ Similarly, writing (34) for $i=1,$ $j=3$
and for $i=3,$ $j=1$ we find $z(\gamma_{1}-\gamma_{3}).$ Then, writing (34)
for $i=2,$ $j=3$ and for $i=3,$ $j=2,$ we find $z(\gamma_{2}-\gamma_{3}).$

\textit{Step 3. }In this step using (28), (35), we find%
\begin{equation}
z(\gamma_{1}+\gamma_{2}-\gamma_{3}),\text{ }z(\gamma_{1}+\gamma_{3}-\gamma
_{2}),\text{ }z(\gamma_{2}+\gamma_{3}-\gamma_{1}).
\end{equation}
Writing (28) and (35) for $i=1,$ and taking into account that $\gamma
=\gamma_{1}+\gamma_{2}+\gamma_{3},$ we get
\begin{equation}
\operatorname{Re}(z(\gamma_{2}+\gamma_{3}-\gamma_{1})z(-\gamma_{2}-\gamma
_{3})z(\gamma_{1}))=c_{6},
\end{equation}%
\begin{equation}
\operatorname{Re}(z^{2}(-\gamma_{1})z(\gamma_{1}+\gamma_{2}+\gamma
_{3})z(\gamma_{1}-\gamma_{2}-\gamma_{3}))=c_{7}.
\end{equation}
Let $z(\gamma_{2}+\gamma_{3}-\gamma_{1})=x+iy.$ Then $z(\gamma_{1}-\gamma
_{2}-\gamma_{3})=x-iy.$ Now using (45), (54) and (55) from (60) and (61), we
obtain the equations
\begin{align}
a_{4}x+b_{4}y  &  =c_{8},\\
\text{ }a_{7}x+b_{7}y  &  =c_{9}.
\end{align}
Since $a_{4}b_{7}-b_{4}a_{7}\neq0,$ due to (46.1), from (62) and (63) we find
$x$ and $y$ and hence

$z(\gamma_{2}+\gamma_{3}-\gamma_{1}).$ In the same way, namely writing (28),
(35) for $i=2$ and for $i=3,$ we find $z(\gamma_{1}+\gamma_{3}-\gamma_{2})$
and $z(\gamma_{1}+\gamma_{2}-\gamma_{1}).$
\end{proof}

\section{On the Stability of the Algorithm}

We determine constructively the potential from the band functions in two
steps. In the first step we have determined the invariants from the band
functions in the paper [10]. In the second step we found the potential from
the invariants in Section 3 of this paper. In this section we consider the
stability of the problems studied in both steps.

First, using the asymptotic formulas (13), (19) and (4) of the paper [10],
denoted here as (13[10]), (19[10]) and (4[10]), we consider the stability of
the invariants (15)-(17) with respect to the errors in the Bloch eigenvalues
for the potential of the form (3). For this let us recall the formulas of [10]
that will be used here.\ In [10] the spectral invariants are expressed by the
band functions of the Schr\"{o}dinger operator $L(q^{\delta})$ with the
directional potential $q^{\delta}(x)$ (see (11)), where $\delta$ is a visible
element of $\Gamma$. The function $q^{\delta}$ depends only on one variable
$s=\langle\delta,x\rangle$ and can be written as
\begin{equation}
q^{\delta}(x)=Q^{\delta}(\langle\delta,x\rangle),\text{ where }Q^{\delta
}(s)=\sum_{n\in\mathbb{Z}}z(n\delta)e^{ins},
\end{equation}
that is, $Q^{a}(s)$ is obtained from the right-hand side of (11) by replacing
$\langle a,x\rangle$ with $s.$ The band functions and the Bloch functions of
the operator $L(q^{\delta})$ are
\[
\lambda_{j,\beta}(v,\tau)=\mid\beta+\tau\mid^{2}+\mu_{j}(v),\text{ }%
\Phi_{j,\beta}(x)=e^{i\langle\beta+\tau,x\rangle}\varphi_{j,v}(s),
\]
where $\beta\in\Gamma_{\delta},$ $\tau\in F_{\delta}=:H_{\delta}%
/\Gamma_{\delta},$ $j\in\mathbb{Z}$, $v\in\lbrack0,1),$ $\mu_{j}(v)$ and
$\varphi_{j,v}(s)\mathbb{\ }$are the eigenvalues and eigenfunctions of the
operator $T_{v}(Q^{\delta})$ generated by the boundary value problem:%
\[
-\mid\delta\mid^{2}y^{\prime\prime}(s)+Q^{\delta}(s)y(s)=\mu y(s),\text{
}y(2\pi)=e^{i2\pi v}y(0),\text{ }y^{^{\prime}}(2\pi)=e^{i2\pi v}y^{^{\prime}%
}(0).
\]
In the paper [10] we constructed a set of eigenvalue, denoted by
$\Lambda_{j,\beta}(v,\tau),$ of $L_{t}(q)$ satisfying%
\begin{equation}
\Lambda_{j,\beta}(v,\tau)=\left\vert \beta+\tau\right\vert ^{2}+\mu
_{j}(v)+\frac{1}{4}\int_{F}\left\vert f_{\delta,\beta+\tau}\right\vert
^{2}\left\vert \varphi_{j,v}\right\vert ^{2}dx+O(\rho^{-3a+2\alpha_{1}}\ln
\rho), \tag{13[10]}%
\end{equation}
where $\beta\sim\rho,$ $j=O(\rho^{\alpha_{1}}),$ $\alpha_{1}=3\alpha,$
$a=406\alpha,$ $\alpha=\frac{1}{432},$ $-3a+2\alpha_{1}=-\frac{101}{36}$ and%
\begin{equation}
f_{\delta,\beta+\tau}(x)=\sum_{\gamma:\gamma\in Q(1,1,1)\backslash
\delta\mathbb{R}\text{ }}\frac{\gamma}{\langle\beta+\tau,\gamma\rangle
}z(\gamma)e^{i\langle\gamma,x\rangle}.
\end{equation}
We say that $a(\rho)$ is of order $b(\rho)$ and write $a(\rho)\sim b(\rho)$ if
there exist positive constants $c_{1}$ and $c_{2}$ such that $c_{1}\mid
b(\rho)\mid<\mid a(\rho)\mid<c_{2}\mid b(\rho)\mid$ for $\ \rho\gg1.$ To
consider the stability of the invariants (15)-(17) with respect to the errors
in the band functions, we use (13[10]) and the following asymptotic
decomposition of $\mu_{j}(v)$ and $\left\vert \varphi_{j,v}(s)\right\vert
^{2}:$
\begin{equation}
\mu_{j}(v)=\mid j\delta\mid^{2}+\frac{c_{1}}{j}+\frac{c_{1}}{j^{2}}%
+...+\frac{c_{n}}{j^{n}}+O(\frac{1}{j^{n+1}}), \tag{AD1}%
\end{equation}
\begin{equation}
\left\vert \varphi_{j,v}(s)\right\vert ^{2}=A_{0}+\frac{A_{1}(s)}{j}%
+\frac{A_{2}(s)}{j^{2}}+...+\frac{A_{n}(s)}{j^{n}}+O(\frac{1}{j^{n+1}}),
\tag{AD2}%
\end{equation}
where%
\begin{equation}
c_{1}=c_{2}=0,\text{ }c_{3}=\frac{1}{16\pi\mid\delta\mid^{3}}\int_{0}^{2\pi
}\left\vert Q^{\delta}(t)\right\vert ^{2}dt
\end{equation}
(see [7] and [1]). In [10] we proved that if $q^{\delta}(x)$ has the form
(12), then
\begin{align}
A_{0}  &  =1,\text{ }A_{1}=0,\text{ }A_{2}=\frac{Q^{\delta}(s)}{2}%
+a_{1}\left\vert z(\delta)\right\vert ^{2},\text{ }A_{3}=a_{2}Q^{\delta
}(s)+a_{3}\left\vert z(\delta)\right\vert ^{2},\tag{19[10]}\\
A_{4}  &  =a_{4}Q^{\delta}(s)+a_{5}((z(\delta))^{2}e^{i2\langle\delta
,x\rangle}+(z(-\delta))^{2}e^{-i2\langle\delta,x\rangle})+a_{6},\nonumber
\end{align}
where $a_{1},a_{2},...,a_{6}$ are the known constants.

\begin{theorem}
Let $q(x)$ be the potential of the form (3), satisfying (40). If the band
functions of order $\rho^{2}$ of $L(q)$ are given with accuracy $O(\rho
^{-\frac{101}{36}}\ln\rho)$ , then one can determine the spectral invariants
(15)-(17), constructively and uniquely, with accuracy $O(\rho^{-\frac{97}%
{108}}\ln\rho)$.
\end{theorem}

\begin{proof}
First, using the asymptotic formula (13[10]), we write explicitly the
asymptotic expression of the invariants
\begin{equation}
\mu_{j}(v)\text{, \ }J(\delta,b,j,v)=\int_{F}\mid q_{\delta,b}(x)\varphi
_{j,v}(\langle\delta,x\rangle)\mid^{2}dx \tag{4[10]}%
\end{equation}
determined constructively in [10], where $\upsilon\in(0,\frac{1}{2})\cup
(\frac{1}{2},1)$, $j\in\mathbb{Z}$, $q_{\delta,b}(x)$ is defined in (14),
$\delta\in Q(1,1,1)$ and $b$ is a visible element of $\Gamma_{\delta},$ in
terms of the band functions with an estimate of the remainder term. Let
$s_{1}b_{1}$, $s_{2}b_{2},...,s_{m}b_{m}$ be projections of the vectors of the
set $Q(1,1,1)\backslash\delta\mathbb{R}$ onto the plane $H_{\delta},$ where
$s_{i}\in\mathbb{R}$ and $b_{i}\in\Gamma_{\delta}$ (see 24). If $b_{i}\in
b_{j\text{ }}\mathbb{R}$, where $i>j,$ then we do not include $b_{i}$ to the
list of projections, that is, $b_{1}$, $b_{2},...,b_{m}$ are pairwise linearly
independent. Consider the planes $P(\delta,b_{k})$ for $k=1,2,...,m$. It is
clear that the set $Q(1,1,1)\backslash\delta\mathbb{R}$ is the union of the
pairwise disjoint sets $P(\delta,b_{k})\cap(Q\backslash\delta\mathbb{R)}$ for
$k=1,2,...,m$.\ \ To find the spectral invariants (4[10]), we write
$f_{\delta,\beta+\tau}(x)$ (see (65)) in the form%
\begin{equation}
f_{\delta,\beta+\tau}(x)=\sum_{k=1}^{m}\text{ }F_{\delta,b_{k},\beta+\tau}(x),
\end{equation}
where
\begin{equation}
F_{\delta,b_{k},\beta+\tau}(x)=\sum_{\gamma:\gamma\in P(\delta,b_{k}%
)\cap(Q\backslash\delta\mathbb{R)}\text{ }}\frac{\gamma}{\langle\beta
+\tau,\gamma\rangle}z(\gamma)e^{i\langle\gamma,x\rangle}.
\end{equation}
Clearly, if $\gamma\in P(\delta,b_{k})\backslash\delta\mathbb{R}$ and
$\gamma^{^{\prime}}\in P(\delta,b_{l})\backslash\delta\mathbb{R}$ for $l\neq
k,$ then $\gamma^{^{\prime}}+\gamma\notin\delta\mathbb{R}$. Therefore taking
into account that $\varphi_{j,v}(\langle\delta,x\rangle)$ is a function of
$\langle\delta,x\rangle,$ we obtain
\[
\text{ }\int_{F}\left\langle \text{ }F_{\delta,b_{k},\beta+\tau}(x),\text{
}F_{\delta,b_{l},\beta+\tau}(x)\right\rangle \mid\varphi_{j,v}(\langle
\delta,x\rangle)\mid^{2}dx=0,\text{ }\forall l\neq k.
\]
This with (67) implies that
\begin{equation}
\int_{F}\mid f_{\delta,\beta+\tau}\mid^{2}\left\vert \varphi_{j,v}\right\vert
^{2}dx=\sum_{k=1}^{m}\text{ }\int_{F}\mid F_{\delta,b_{k},\beta+\tau}\mid
^{2}\left\vert \varphi_{j,v}\right\vert ^{2}dx
\end{equation}
In [10] (see (58) of [10]) we proved that for each $b_{0}\in\Gamma_{\delta}$
there exists $\beta_{0}+\tau$ such that
\[
\mid\beta_{0}+\tau\mid\sim\rho,\text{ }\frac{1}{3}\rho^{a}<\mid\langle
\beta_{0}+\tau,b_{0}\rangle\mid<3\rho^{a},
\]
and $\Lambda_{j,\beta_{0}}(v,\tau)$ satisfies (13[10]). Since $b_{k}\in
\Gamma_{\delta},$ there exist $\beta_{k}+\tau$ such that
\begin{equation}
\frac{1}{3}\rho^{a}<\mid\langle\beta_{k}+\tau,b_{k}\rangle\mid<3\rho^{a}%
\end{equation}
and $\Lambda_{j,\beta_{0}}(v,\tau)$ satisfies (13[10]). From (70) we see that
$\cos\theta_{k,k}=O(\rho^{a-1})=o(1),$ where $\theta_{s,k}$ is the angle
between the vectors $\beta_{s}+\tau$ and $b_{k}$. Therefore $\cos\theta
_{s,k}\sim1$ for $s\neq k$ and hence
\begin{equation}
\langle\beta_{s}+\tau,b_{k}\rangle\sim\rho
\end{equation}
for all $s\neq k.$ If $b_{0}\notin b_{1}\mathbb{R}\cup b_{2}\mathbb{R}%
\cup...\cup b_{m}\mathbb{R}$, then (71) holds for $k=0$ and $s=1,2,...,m$.

Now substituting the orthogonal decomposition $|\delta|^{-2}\langle
\gamma,\delta\rangle\delta+|b_{k}|^{-2}\langle\gamma,b_{k}\rangle b_{k}$ of
$\gamma$ for $\gamma\in P(\delta,b_{k})\cap(Q\backslash\delta\mathbb{R)}$ into
the denominator of the fraction in (68), and taking into account that
$\beta+\tau\in H_{\delta},$ $\langle\beta+\tau,\delta\rangle=0,$ we obtain%
\[
F_{\delta,b_{k},\beta+\tau}(x)=\frac{|b_{k}|^{2}}{\langle\beta+\tau
,b_{k}\rangle}q_{\delta,b_{k}}(x),
\]
where $q_{\delta,b_{k}}(x)$ is defined in (14). This with (4[10]) implies
that
\begin{equation}
\int_{F}\mid F_{\delta,b_{k},\beta+\tau}\mid^{2}\left\vert \varphi
_{j,v}\right\vert ^{2}dx=\frac{|b_{k}|^{4}}{(\langle\beta+\tau,b_{k}%
\rangle)^{2}}J(\delta,b_{k},j,v).
\end{equation}

Substituting (69) and (72) in (13[10]) and then instead of $\beta$ writing
$\beta_{s}$ for $s=0,1,...,m,$ we get the system of $m+1$ equations
\begin{equation}
\mu_{j}(v)+\sum_{k=1}^{m}\frac{|b_{k}|^{4}}{4(\langle\beta_{s}+\tau
,b_{k}\rangle)^{2}}J(\delta,b_{k},j,v)=\Lambda_{j,\beta_{s}}(v,\tau
)+\left\vert \beta_{s}+\tau\right\vert ^{2}+O(\rho^{-3a+2\alpha_{1}}\ln\rho),
\end{equation}
with respect to the unknowns $\mu_{j}(v),J(\delta,b_{1},j,v),$ $J(\delta
,b_{2},j,v),...,J(\delta,b_{m},j,v).$ By (70) and (71) the coefficient matrix
of (73) is $(a_{i,j})$, where $a_{i1}=1$ for $i=1,2,...,m+1$ and
\begin{equation}
a_{k,k}\sim\rho^{-2a},\text{ }a_{s,k}\sim\rho^{-2},\text{ }\forall k>1,\text{
}\forall s\neq k.
\end{equation}
Expanding the determinant $\Delta$ of the matrix $(a_{i,j}),$ one can readily
see that the highest order term of this expansion is the product of the
diagonal elements of the matrix $(a_{i,j})$ which is of order $\rho^{-2ma}$
and the other terms of this expansions are $O(\rho^{-2m}).$ Therefore, we
have
\begin{equation}
\Delta\sim\rho^{-2ma}%
\end{equation}
Now we are going to use the fact that the right-hand side of (73) is
determined with error $O(\rho^{-3a+2\alpha_{1}}\ln\rho),$ if the band
functions of order $\rho^{2}$ of $L(q)$ are given with accuracy $O(\rho
^{-3a+2\alpha_{1}}\ln\rho).$\ Let $\Delta_{k},$ $\Delta_{k,0}$ and
$\Delta_{k,1}$ be determinant obtained from $\Delta$ by replacing $s$-th
elements of the $k$-th column by
\[
\Lambda_{j,\beta_{s}}(v,\tau)+\mid\beta_{s}+\tau\mid^{2}+O(\rho^{-3a+2\alpha
_{1}}\ln\rho),\text{ }\Lambda_{j,\beta_{s}}(v,\tau)+\mid\beta_{s}+\tau\mid
^{2})
\]
and $O(\rho^{-3a+2\alpha_{1}}\ln\rho)$ respectively. One can easily see that%
\begin{equation}
\Delta_{1}-\Delta_{1,0}=\Delta_{1,1}=O(\rho^{-2ma-3a+2\alpha_{1}}\ln
\rho),\text{ }\Delta_{k}-\Delta_{k,0}=\Delta_{k,1}=O(\rho^{-2ma-a+2\alpha_{1}%
}\ln\rho)
\end{equation}
for $k>1.$ Therefore, solving the system (73) by the Cramer's rule and using
(75), (76), we find $\mu_{j}(v)$ and $J(\delta,b_{k},j,v)$ with error
$O(\rho^{-3a+2\alpha_{1}}\ln\rho)$ and $O(\rho^{-a+2\alpha_{1}}\ln\rho)$ respectively.

Now using (AD1) for $j\sim\rho^{\alpha_{1}}$, where $n$ is chosen so that
$j^{n+1}>\rho^{3a},$ and taking into account that $\mu_{j}(v)$ is determined
with error $O(\rho^{-3a+2\alpha_{1}}\ln\rho)$, we consider the invariant (15).
In (AD1) replacing $j$ by $kj,$ for $k=1,2,...,n,$ we get the system of $n$
equations
\begin{equation}
\frac{c_{1}}{jk}+\frac{c_{2}}{(jk)^{2}}+...+\frac{c_{n}}{(jk)^{n}}=\mu
_{jk}(v)+\mid jk\delta\mid^{2}+O(\frac{1}{j^{n+1}}),
\end{equation}
with respect to the unknowns $c_{1},c_{2},...,c_{n}.$ The coefficient matrix
of this system is $(a_{i,k})$, where $a_{i,k}=\frac{c_{k}}{(ji)^{k}}$ for
$i,k=1,2,...,n$. Therefore the determinant of $(a_{i,k})$ is
\[
\frac{c_{1}}{j}\frac{c_{2}}{j^{2}}...\frac{c_{n}}{j^{n}}\det(v_{i,k}),
\]
where $v_{i,k}=v_{i}^{k},v_{i}=\frac{1}{i},$ that is, $(v_{i,k})$ is the
Vandermonde matrix and $\det(v_{i,k})\sim1.$ Now solving the system (77) by
the Cramer's rule and using the arguments used for the solving of (73), we
find $c_{3}$ with an accuracy $O(\rho^{-3a+5\alpha_{1}}\ln\rho),$ since the
elements of the third column is of order $\rho^{3\alpha_{1}}$ and the
right-hand side of (77) is determined with error $O(\rho^{-3a+2\alpha_{1}}%
\ln\rho).$ Thus formula (66) gives the invariant (15) with error
$O(\rho^{-3a+5\alpha_{1}}\ln\rho).$

To consider the invariant (16) and (17), we use (AD2), where $j\sim
\rho^{\alpha_{1}}$ and $n$ can be chosen so that $j^{n+1}>\rho^{a}.$ In (AD2)
replacing $j$ by $kj,$ for $k=1,2,...,n+1,$ and using it in $J(\delta
,b_{s},j,v)$ (see (4[10])), we get the system of $n+1$ equations
\begin{equation}
J_{0}(\delta,b_{s})+\frac{J_{1}(\delta,b_{s})}{jk}+\frac{J_{2}(\delta,b_{s}%
)}{(jk)^{2}}+...+\frac{J_{n}(\delta,b_{s})}{(jk)^{n}}=J(\delta,b_{s},j,v),
\end{equation}
with respect to the unknowns $J_{0}(\delta,b_{s}),J_{1}(\delta,b_{s}%
),...,J_{n}(\delta,b_{s}),$ where $\ $%
\[
J_{k}(\delta,b_{s})=\int_{F}|q_{\delta,b_{s}}(x)|^{2}A_{k}(\langle
\delta,x\rangle)dx.
\]
In the above we proved that the write-hand side of (78) is determined with
error $O(\rho^{-a+2\alpha_{1}}\ln\rho).$ Therefore, instead of (77) using (78)
and repeating the arguments used in the finding of $c_{3}$, we find
$J_{0}(\delta,b_{s}),J_{1}(\delta,b_{s}),...,J_{4}(\delta,b_{s})$ with
accuracy $O(\rho^{-a+6\alpha_{1}}\ln\rho)$. Then using (19 [10]), we determine
the invariants (16) and (17) with the accuracy $O(\rho^{-a+6\alpha_{1}}\ln
\rho),$ where $a-6\alpha_{1}=\frac{97}{108}$
\end{proof}

Now considering the proof of Theorem 4, we will show that if the invariants
are given with error $\varepsilon,$ where $\varepsilon\ll1,$ then the Fourier
coefficients can be determined with error $\varepsilon.$ For this we use the
following simplest lemma.

\begin{lemma}
Let $x(\varepsilon)$ and $y(\varepsilon)$ be the solution of the system of the
equations%
\[
(a+\varepsilon)x+(b+\varepsilon)y=e+\varepsilon,\text{ }(c+\varepsilon
)x+(d+\varepsilon)y=f+\varepsilon.
\]
If $ad-cb\neq0,$ then $x(\varepsilon)=x(0)+O(\varepsilon)$ and $y(\varepsilon
)=y(0)+O(\varepsilon).$
\end{lemma}

\begin{proof}
Solving the system of equations by Cramer's rule we get
\[
x(\varepsilon)=\frac{(e+\varepsilon)(d+\varepsilon)-(b+\varepsilon
)(f+\varepsilon)}{(a+\varepsilon)(d+\varepsilon)-(b+\varepsilon)(c+\varepsilon
)}.
\]
Since $(a+\varepsilon)(d+\varepsilon)-(b+\varepsilon)(c+\varepsilon
)=ad-bc+O(\varepsilon),$ $ad-bc\neq0$ and

$(e+\varepsilon)(d+\varepsilon)-(b+\varepsilon)(f+\varepsilon
)=ed-bf+O(\varepsilon)$ we have $x(\varepsilon)=x(0)+O(\varepsilon).$ In the
same way we get $y(\varepsilon)=y(0)+O(\varepsilon)$
\end{proof}

\begin{theorem}
Let $q(x)$ be the potential of the form (3) satisfying (40). If the spectral
invariants (15)-(17) are given with error $\varepsilon,$ then the potential
$q$ can be determined constructively and uniquely, modulo (2), with error
$O(\varepsilon),$ where $\varepsilon$ is a small number.
\end{theorem}

\begin{proof}
$(a)$ If the invariant (15) is given with the error $\varepsilon$, then by
using (20) and (44) we determine the Fourier coefficients $z(\gamma
_{1}),z(\gamma_{2})$ and $z(\gamma_{3})$ with error $O(\varepsilon).$ It
follows from the proof of (49) that if the invariants (25) and (20) are given
with the error $O(\varepsilon)$, then one can determine the real parts and the
absolute values of the imaginary parts of the Fourier coefficients
$z(\gamma_{2}+\gamma_{3}),z(\gamma_{1}+\gamma_{3})$ and $z(\gamma_{2}%
+\gamma_{3})$ with error $O(\varepsilon).$ One can readily see from the proof
of \textit{Step 1 }of\textit{ }Theorem 4 that the error \ $O(\varepsilon)$ in
the $a_{m}$ and $\mid b_{m}\mid$ for $m=4,5,6$ does not influence the
determinations of the signs of $t_{4},t_{5}$ and $t_{6}.$ Therefore the
Fourier coefficients $z(\gamma_{2}+\gamma_{3}),z(\gamma_{1}+\gamma_{3})$ and
$z(\gamma_{2}+\gamma_{3})$ can be determined with error $O(\varepsilon).$ The
Fourier coefficients in (55), (56) and (59) were determined from the systems
of equations generated by pairs \{(50), (51)\}, \{(60), (61)\} and \{(62),
(63)\}. Moreover, by (46.1), the main determinants $a_{4}b_{5}-b_{4}a_{5},$
$2a_{6}b_{6}$ and $a_{4}b_{7}-b_{4}a_{7}$ of these systems are not zero. Thus
Lemma 1 implies that if the invariants are given with the error $O(\varepsilon
)$, then the Fourier coefficient in (55), (56) and (59) can be determined with
error $O(\varepsilon).$
\end{proof}

The consequence of Theorem 5 and Theorem 6 is the following:

\begin{corollary}
Let $q(x)$ be the potential of the form (3) satisfying (40). If the band
functions of order $\rho^{2}$ of $L(q)$ are given with accuracy $O(\rho
^{-\frac{101}{36}}\ln\rho)$, then one can determine the potential $q$
constructively and uniquely, modulo (2), with accuracy $O(\rho^{-\frac
{97}{108}}\ln\rho)$.
\end{corollary}

\section{Uniqueness Theorems}

First we consider the Hill operator $H(p)$ generated in $L_{2}(\mathbb{R})$ by
the expression

$l(q)=:-y^{^{\prime\prime}}(x)+p(x)y(x),$ when $p(x)$ is a real-valued
trigonometric polynomial%
\begin{equation}
p(x)=%
%TCIMACRO{\tsum \limits_{s=-N}^{N}}%
%BeginExpansion
{\textstyle\sum\limits_{s=-N}^{N}}
%EndExpansion
p_{s}e^{2isx},\text{ }p_{-s}=\overline{p_{s}},\text{ }p_{0}=0.
\end{equation}
Let the pair $\{\lambda_{k,1}$, $\lambda_{k,2}\}$ denote, respectively, the
$k$-th eigenvalues of the operator generated in $L_{2}[0,\pi]$ by the
expression $l(q)$ and the periodic boundary conditions for $k$ even and the
anti-periodic boundary conditions for $k$ odd. It is well-known that (see [1],
Theorem 4.2.4)
\[
\lambda_{0,1}=\lambda_{0,2}<\lambda_{1,1}\leq\lambda_{1,2}<\lambda_{2,1}%
\leq\lambda_{2,2}<\lambda_{3,1}\leq\lambda_{3,2}<...<\lambda_{n,1}\leq
\lambda_{n,2}<...\text{ .}%
\]
\ The spectrum $Spec(H(p))$ of $H(p)$ is the union of the intervals
$[\lambda_{n-1,2},\lambda_{n,1}]$ for $n=1,2,...$. The interval $\gamma
_{n}=:(\lambda_{n,1},\lambda_{n,2})$ is the $n$-th gaps in the spectrum of
$H(p).$ Since the spectrum of the operators $H(p(x))$ and $(H(p(x+\tau))$,
where $\tau\in(0,\pi),$ are the same, we may assume, without loss of
generality, that $p_{-N}=p_{N}=\mu>0.$ We use the following formula obtained
in the paper [5] ( see Theorem 2 in [5]) for the length $\mid\gamma_{n}\mid$
of the gap $\gamma_{n}:$%
\begin{equation}
\mid\gamma_{n}\mid=\frac{4n}{\mu}\left(  \frac{\mu e^{2}}{8n^{2}}\right)
^{\frac{n}{N}}\left\vert
%TCIMACRO{\tsum \limits_{k=0}^{N-1}}%
%BeginExpansion
{\textstyle\sum\limits_{k=0}^{N-1}}
%EndExpansion
A_{k}(n)\left(  1+O\left(  \frac{\ln n}{n}\right)  \right)  \right\vert ,
\end{equation}
where
\begin{equation}
A_{k}(n)=\exp\left[  \frac{2ink\pi}{N}+2n%
%TCIMACRO{\tsum \limits_{j=1}^{N-1}}%
%BeginExpansion
{\textstyle\sum\limits_{j=1}^{N-1}}
%EndExpansion
\lambda_{j}\left(  \left(  \tfrac{1}{2}\mu n^{-2}\right)  ^{\frac{1}{N}%
}e^{2ik\pi/N}\right)  ^{j}\right]
\end{equation}
and $\lambda_{j}$ algebraically depends on the Fourier coefficients of $p(x).$

From (81) one can readily see that%
\begin{equation}
\mid A_{k}(n)\mid<\exp(an^{1-\frac{2}{N}}),\text{ }\mid A_{k}(n)\mid
>\exp(-an^{1-\frac{2}{N}}),\text{ }\forall k=0,1,...,(N-1),
\end{equation}
where
\begin{equation}
a=%
%TCIMACRO{\tsum \limits_{j=1}^{N-1}}%
%BeginExpansion
{\textstyle\sum\limits_{j=1}^{N-1}}
%EndExpansion
a_{j},\text{ }a_{j}=\sup_{k}\left\vert \operatorname{Re}(2\lambda_{j}\left(
\left(  \tfrac{1}{2}\mu\right)  ^{\frac{1}{N}}e^{2ik\pi/N}\right)
^{j}\right\vert .
\end{equation}
This and (80) imply that%
\begin{equation}
\mid\gamma_{n}\mid<\frac{4n}{\mu}\left(  \frac{\mu e^{2}}{8n^{2}}\right)
^{\frac{n}{N}}2Ne^{an^{1-\frac{2}{N}}}.
\end{equation}

Using (82)-(84) we prove the following:

\begin{theorem}
Let $\widetilde{p}(x)$ be a real-valued trigonometric polynomial of the form
\[
\widetilde{p}(x)=%
%TCIMACRO{\tsum \limits_{s=-K}^{K}}%
%BeginExpansion
{\textstyle\sum\limits_{s=-K}^{K}}
%EndExpansion
\widetilde{p}_{s}e^{2isx},\text{ }\widetilde{p}_{-s}=\overline{\widetilde
{p}_{s}},\text{ }\widetilde{p}_{-K}=\widetilde{p}_{K}=\nu>0.
\]
If $Spec(H(p))=Spec(H(\widetilde{p})),$ then $K=N,$ where $p(x)$ is defined in (64).
\end{theorem}

\begin{proof}
Suppose $K\neq N$. Without less of generality, it can be assumed that $K<N.$
We consider the following two cases:

Case 1: Assume that $\lambda_{j}=0$ for all values of $j.$ Then by (80) for
$n=lN$ and for $l\gg1$ we have $A_{k}(n)=1$ for all $k$. Therefore, by (80),
we have%
\begin{equation}
\mid\gamma_{n}\mid=\frac{4n}{\mu}\left(  \frac{\mu e^{2}}{8n^{2}}\right)
^{l}N\left(  1+O\left(  \frac{\ln n}{n}\right)  \right)  ,\forall n=lN.
\end{equation}
Applying (84) for the length $\mid\delta_{n}\mid$ of the $n$ th gap
$\delta_{n}$ in the $Spec(H(\widetilde{p})),$ that is, replacing $N$ and $\mu$
by $K$ and $\nu$ respectively and arguing as in the proof of (84), we see that
there exist a positive number $b$ such that%
\begin{equation}
\mid\delta_{n}\mid<\frac{4n}{\nu}\left(  \frac{\nu e^{2}}{8n^{2}}\right)
^{\frac{n}{K}}2Ke^{bn^{1-\frac{2}{K}}}.
\end{equation}
Since the fastest decreasing multiplicands of (85) and (86) are $n^{-2l}$ and
$n^{-\frac{2n}{K}}$ respectively and $\ K<N,$ it follows from (85) and (86)
for $n=lN$ that $\mid\gamma_{lN}\mid>\mid\delta_{lN}\mid$ for $l\gg1,$ which
contradicts to the equality $Spec(H(q))=Spec(H(\widetilde{p})).$

Case 2: Assume that $\lambda_{j}\neq0$ for some values of $j.$ Let us prove
that the equalities
\begin{equation}
\mid\gamma_{lN}\mid=\mid\delta_{lN}\mid,\mid\gamma_{lN+1}\mid=\mid
\delta_{lN+1}\mid,...,\mid\gamma_{lN+N-1}\mid=\mid\delta_{lN+N-1}\mid
\end{equation}
for $l\gg1$ can not be satisfied simultaneously. Suppose to the contrary that
all equalities in (87) hold. Using (80), (86) and taking into account that%
\[
\left(  \frac{\nu e^{2}}{8n^{2}}\right)  ^{\frac{lN+m}{K}}\left(  \frac{\mu
e^{2}}{8n^{2}}\right)  ^{-\frac{lN+m}{N}}e^{bn^{1-\frac{2}{K}}}=O(n^{-\alpha
n})
\]
for $0<\alpha<\frac{lN+m}{K}-\frac{lN+m}{N},$ from (87) we obtain%
\begin{equation}%
%TCIMACRO{\tsum \limits_{k=0}^{N-1}}%
%BeginExpansion
{\textstyle\sum\limits_{k=0}^{N-1}}
%EndExpansion
A_{k}(lN+m)\left(  1+O\left(  \frac{\ln l}{l}\right)  \right)  =O(l^{-\alpha
l}),\text{ }\forall m=0,1,...(N-1).
\end{equation}
Let us consider $A_{k}(lN+m)$ \ in detail. It can be written in the form%
\begin{equation}
A_{k}(lN+m)=\exp\left(  \frac{2imk\pi}{N}\right)  e^{c_{k}(lN+m)},\text{
}c_{k}(lN+m)=%
%TCIMACRO{\tsum \limits_{j=1}^{N-1}}%
%BeginExpansion
{\textstyle\sum\limits_{j=1}^{N-1}}
%EndExpansion
M_{j}(k)(lN+m)^{1-\frac{2j}{N}},
\end{equation}
where $M_{j}(k)$ is a complex number. Using the mean value theorem, we get%
\begin{equation}
c_{k}(lN+m)-c_{k}(lN)=m%
%TCIMACRO{\tsum \limits_{j=1}^{N-1}}%
%BeginExpansion
{\textstyle\sum\limits_{j=1}^{N-1}}
%EndExpansion
M_{j}(k)(lN+\theta(k))^{-\frac{2j}{N}}=O(l^{-\frac{2}{N}}),
\end{equation}
where $\theta(k)\in\lbrack0,m]$ for all $k.$ Now using (89), (90) and taking
into account that

$e^{z}=1+O(z)$ as $z\rightarrow0,$ we obtain%
\begin{equation}
A_{k}(lN+m)=\exp\left(  \frac{2imk\pi}{N}\right)  A_{k}(lN)(1+O(l^{-\frac
{2}{N}})).
\end{equation}
Therefore (88) has the form
\begin{equation}%
%TCIMACRO{\tsum \limits_{k=0}^{N-1}}%
%BeginExpansion
{\textstyle\sum\limits_{k=0}^{N-1}}
%EndExpansion
\exp\left(  \frac{2imk\pi}{N}\right)  A_{k}(lN)\left(  1+o(1)\right)
=O(l^{-\alpha l}),\text{ }m=0,1,...(N-1).
\end{equation}
Consider (92) as a system of equations with respect to the unknowns

$A_{0}(lN),$ $A_{1}(lN),...,$ $A_{N-1}(lN).$ Using the well-known formula for
the determinant of the Vandermonde matrix $(v_{m,k})$, where $v_{m,k}%
=v_{m}^{k},v_{m}=\exp(\frac{2im\pi}{N}),$ we see that the main determinant of
this system is
\[
\left(  1+o(1)\right)  \det\left(  e^{\frac{2imk\pi}{N}}\right)
_{k,m=0}^{N-1}=\left(  1+o(1)\right)
%TCIMACRO{\tprod \limits_{0\leq m<k\leq(N-1)}}%
%BeginExpansion
{\textstyle\prod\limits_{0\leq m<k\leq(N-1)}}
%EndExpansion
(e^{\frac{2ik\pi}{N}}-e^{\frac{2im\pi}{N}}).
\]
Thus solving (92) by the Cramer's rule we obtain $A_{k}(lN)=O(l^{-\alpha l}),$
for $k=0,1,...(N-1)$ which contradicts the second inequality in (82). The
theorem is proved.
\end{proof}

Now using this theorem we prove a uniqueness theorem for the three-dimensional
Schr\"{o}dinger operator. For this, first, we prove the following lemma.

\begin{lemma}
Let $\widetilde{q}(x)$ be infinitely differentiable$\ $periodic potential of
the form%
\begin{equation}
\widetilde{q}(x)=%
%TCIMACRO{\tsum \limits_{a\in Q(1,1,1)}}%
%BeginExpansion
{\textstyle\sum\limits_{a\in Q(1,1,1)}}
%EndExpansion
\widetilde{q}^{a}(x),
\end{equation}
where
\begin{equation}
\widetilde{q}^{a}(x)=%
%TCIMACRO{\tsum \limits_{n\in\mathbb{Z}}}%
%BeginExpansion
{\textstyle\sum\limits_{n\in\mathbb{Z}}}
%EndExpansion
\widetilde{z}(na)e^{in\langle a,x\rangle},\text{ }\widetilde{z}(0)=0
\end{equation}
and $\widetilde{z}(na)=:(\widetilde{q}(x),e^{in\langle a,x\rangle})$ is the
Fourier coefficients of $\widetilde{q}$. If the equalities
\begin{equation}
\widetilde{z}(n\gamma_{i})=0,\text{ }\widetilde{z}(n\gamma_{j})=0,\text{
}\forall n\in\mathbb{Z}\backslash\{-1,1\}
\end{equation}
hold, then
\begin{equation}
\widetilde{I}_{1}(\gamma_{i}+\gamma_{j},\ \gamma_{i})=A_{1}(\gamma_{i}%
+\gamma_{j},\ \gamma_{i})\operatorname{Re}(\widetilde{z}(-\gamma_{i}%
-\gamma_{j})\widetilde{z}(\gamma_{i})\widetilde{z}(\gamma_{j})),\text{ }
\tag{$\widetilde{25}$}%
\end{equation}%
\begin{equation}
\widetilde{I}_{1}(\gamma_{i}-\gamma_{j},\ \gamma_{i})=A_{1}(\gamma_{i}%
-\gamma_{j},\ \gamma_{i})\operatorname{Re}(\widetilde{z}(-\gamma_{i}%
+\gamma_{j})\widetilde{z}(\gamma_{i})\widetilde{z}(-\gamma_{j})),\text{ }
\tag{$\widetilde{26}$}%
\end{equation}%
\begin{equation}
\widetilde{I}_{2}(\gamma_{i},\ \gamma_{j})=A_{2}(\gamma_{i},\gamma
_{j})\operatorname{Re}(\widetilde{z}(-\gamma_{i}))^{2}\widetilde{z}(\gamma
_{i}+\gamma_{j})\widetilde{z}(\gamma_{i}-\gamma_{j})) \tag{$\widetilde{34}$}%
\end{equation}
for $i\neq j,$ where $A_{1}(a,b)$ and $A_{2}(a,b)$ are defined in Theorem 1
and Theorem 2 respectively, $\widetilde{I}_{1}(a,b)$ and $\widetilde{I}%
_{2}(a,b)$ are the invariants (16) and (17) for the operator $L(\widetilde
{q})$.
\end{lemma}

\begin{proof}
By definition of $\widetilde{I}_{1}(\gamma_{i}+\gamma_{j},\ \gamma_{i})$ (see
(16) and (14)) we have%
\begin{equation}
\widetilde{I}_{1}(\gamma_{i}+\gamma_{j},\ \gamma_{i})=\int_{F}\left\vert
\widetilde{q}_{\gamma_{i}+\gamma_{j},\beta}(x)\right\vert ^{2}(\widetilde
{q})^{\gamma_{i}+\gamma_{j}}(x)dx,
\end{equation}
where $\beta$ is defined by (24),%
\begin{equation}
\widetilde{q}_{\gamma_{i}+\gamma_{j},\beta}(x)=\sum\limits_{c\in D}\frac
{c}{\langle\beta,c\rangle}\widetilde{z}(c)e^{i\langle c,x\rangle},
\end{equation}
and $D=\{c\in(P(\gamma_{i},\gamma_{j})\cap\Gamma)\backslash(\gamma_{i}%
+\gamma_{j})\mathbb{R}:\widetilde{z}(c)\neq0\}.$ It follows from (93) that if
$c\in D$, then $c=ka,$ where $k$ is an integer, and $a$ belongs to the set
$P(\gamma_{i},\gamma_{j})\cap Q)\backslash(\gamma_{i}+\gamma_{j})\mathbb{R}$.
Since this set is $\{\gamma_{i},\gamma_{j},-\gamma_{i},-\gamma_{j},\gamma
_{i}-\gamma_{j},-(\gamma_{i}-\gamma_{j})\}$ and (95) holds, we have
\begin{equation}
D=\{\gamma_{i},\gamma_{j},-\gamma_{i},-\gamma_{j}\}\cup\{k(\gamma_{i}%
-\gamma_{j}):k\in\mathbb{Z}\}.
\end{equation}
Therefore, repeating the proof of (32), we see that
\begin{equation}
\widetilde{I}_{1}(\gamma_{i}+\gamma_{j},\ \gamma_{i})=2\operatorname{Re}%
\left(  \sum\limits_{n=1}^{\infty}\widetilde{z}(-n(\gamma_{i}+\gamma_{j}%
))\sum\limits_{c\in D}\frac{\left\langle n(\gamma_{i}+\gamma_{j}%
)-c,c\right\rangle }{(\left\langle c,\beta\right\rangle )^{2}}\widetilde
{z}(n(\gamma_{i}+\gamma_{j})-c)\widetilde{z}(c)\right)  .
\end{equation}
It follows from (98) that if $n>1$ and $c\in D,$ then $n(\gamma_{i}+\gamma
_{j})-c\notin D$ and $\widetilde{z}(n(\gamma_{i}+\gamma_{j})-c)=0.$ Hence,
from (99) we obtain
\begin{equation}
\widetilde{I}_{1}(\gamma_{i}+\gamma_{j},\ \gamma_{i})=2\operatorname{Re}%
\left(  \widetilde{z}(-(\gamma_{i}+\gamma_{j}))\sum\limits_{c\in D}%
\frac{\left\langle (\gamma_{i}+\gamma_{j})-c,c\right\rangle }{(\left\langle
c,\beta\right\rangle )^{2}}\widetilde{z}((\gamma_{i}+\gamma_{j})-c)\widetilde
{z}(c)\right)  .
\end{equation}
Using this instead of (32) and repeating the proof of (25), we get
($\widetilde{\text{25}}$). In ($\widetilde{\text{25}}$) replacing $\gamma_{j}$
by $-\gamma_{j},$ we get ($\widetilde{\text{26}}$).

Now let us prove ($\widetilde{\text{34}}$). It follows from (95) that
\[
(\widetilde{q})^{\gamma_{i}}(x)=\widetilde{z}(\gamma_{i})e^{i\langle\gamma
_{i},x\rangle}+\widetilde{z}(-\gamma_{i})e^{-i\langle\gamma_{i},x\rangle}).
\]
Therefore $\widetilde{I}_{2}(\gamma_{i},\gamma_{j})$ has the form%
\begin{equation}
\widetilde{I}_{2}(\gamma_{i},\gamma_{j})=\int_{F}\left\vert \widetilde
{q}_{\gamma_{i},\beta}(x)\right\vert ^{2}(((\widetilde{z}(\gamma_{i}%
))^{2}e^{i2\langle\gamma_{i},x\rangle}+(\widetilde{z}(-\gamma_{i}%
))^{2}e^{-i2\langle\gamma_{i},x\rangle})dx
\end{equation}
(see (17)), where $\beta$ is defined by (24),%
\begin{equation}
\widetilde{q}_{\gamma_{i},\beta}(x)=\sum\limits_{c\in E}\frac{c}{\langle
\beta,c\rangle}\widetilde{z}(c)e^{i\langle c,x\rangle},
\end{equation}
$E=\{c\in(P(\gamma_{i},\gamma_{j})\cap\Gamma)\backslash\gamma_{i}%
\mathbb{R}:\widetilde{z}(c)\neq0\}.$ Arguing as in the proof of (98), (99), we
see that
\begin{equation}
E=\{\gamma_{j},-\gamma_{j}\}\cup\{k(\gamma_{i}-\gamma_{j}):k\in\mathbb{Z}%
\}\cup\{n(\gamma_{i}+\gamma_{j}):n\in\mathbb{Z}\}.
\end{equation}%
\begin{equation}
\widetilde{I}_{2}(\gamma_{i},\gamma_{j})=2\operatorname{Re}\left(
\widetilde{z}^{2}(-\gamma_{i})\sum\limits_{c\in E}\frac{\left\langle
\gamma_{i}+c,\gamma_{i}-c\right\rangle }{(\left\langle c,\beta\right\rangle
)^{2}}\widetilde{z}(\gamma_{i}+c)\widetilde{z}(\gamma_{i}-c)\right)  .
\end{equation}
If $c=k(\gamma_{i}-\gamma_{j}),$ where $k\neq0,$ or $c=n(\gamma_{i}+\gamma
_{j}),$ where $n\neq0,$ then at least one of the vectors $\gamma_{i}-c$ and
$\gamma_{i}+c$ does not have the form $c=sa,$ where $s\in\mathbb{Z}$, $a\in
P(\gamma_{i},\gamma_{j})\cap Q)\backslash\gamma_{i}\mathbb{R}$, and hence by
(93) we have $\widetilde{z}(\gamma_{i}+c)\widetilde{z}(\gamma_{i}-c)=0$.
Therefore, the summation in (104) is taken over $c\in\{\pm\gamma_{j}\}$ and
($\widetilde{\text{34}}$) holds.
\end{proof}

Now we prove a uniqueness theorem for the periodic, with respect to the
lattice $\Omega,$ potentials $q(x)$ of $C^{1}(\mathbb{R}^{3})$ subject to some
constraints only on the directional potentials (see (10), (11)) $q^{\gamma
_{1}}(x),$ $q^{\gamma_{2}}(x)$ and $q^{\gamma_{3}}(x),$ where $\{\gamma
_{1},\gamma_{2},\gamma_{3}\}$ is a basis of $\Gamma$ satisfying (6). Note that
the directional potential $q^{a}(x)$ is a function $Q^{a}(s)$ of one variable
$s=:\langle x,a\rangle\in\mathbb{R}$, where the function $Q^{a}(s)$ is
obtained from the right-hand side of (11) by replacing $\langle x,a\rangle$
with $s,$ that is, $Q^{a}(\langle x,a\rangle)=q^{a}(x)$ (see (64)). Let $M$ be
the set of all periodic,with period $2\pi,$ functions $f\in C^{1}(\mathbb{R})$
such that $spec(H(f))=spec(H(\mu\cos s))$ for some positive $\mu.$ Denote by
$W$ the set of all periodic, with respect to the lattice $\Omega,$ functions
$q(x)$ of $C^{1}(\mathbb{R}^{3})$ whose directional potentials $q^{\gamma_{k}%
}(x)$ for $k=1,2,3$ satisfy the conditions%
\begin{equation}
Q^{\gamma_{k}}\in(C^{1}(\mathbb{R})\backslash M)\cup P,\text{ }\forall
k=1,2,3,
\end{equation}
where $P$ is the set of all trigonometric polynomial. Thus we put condition
only on the directional potentials $q^{\gamma_{1}}(x),$ $q^{\gamma_{2}}(x)$
and $q^{\gamma_{3}}(x).$ The all other directional potentials, that is,
$q^{a}(x)$ for all $a\in S\backslash\{\gamma_{1},\gamma_{2},\gamma_{3}\},$
where $S$ is the set of all visible elements of $\Gamma,$ are arbitrary
continuously differentiable functions.

\begin{theorem}
Let $q(x)$ be the potential of the form (3), satisfying (40). If
$\widetilde{q}\in W$ and the band functions of the operators $L(q)$ and
$L(\widetilde{q})$ coincide, then $\widetilde{q}$ is equal to $q$ modulo (2).
\end{theorem}

\begin{proof}
Let $\widetilde{q}$ be a function of $W$ whose band functions coincides with
the band functions of $q.$ By Theorem 6.1 of [2] the band functions of
$L(\widetilde{q}^{a})$ coincides with the band functions of $L(q^{a}).$ It
implies that the spectrum of $H(\widetilde{Q}^{a})$ coincides with the
spectrum of $H(Q^{a}),$ where $\widetilde{Q}^{a}(\langle x,a\rangle
)=\widetilde{q}^{a}(x).$ Since the length of the $n$-th gap in the spectrum of
$H(Q^{a})$ satisfies (84), the same formula holds for the $n$-th gap of
$H(\widetilde{Q}^{a}).$ It implies that $\widetilde{q}^{a}$ is an infinitely
differentiable function for all visible elements $a$ of $\Gamma$ (see [7]).
Thus $\widetilde{q}(x)$ is an infinitely differentiable function and due to
[10] the operator $L(\widetilde{q})$ has the invariants (15)-(17)$\ $denoted
by $\widetilde{I}(a),$ $\widetilde{I}_{1}(a,b),$ $\widetilde{I}_{2}(a,b).$
Since the band functions of $L(q)$ and $L(\widetilde{q})$ coincide, we have
\begin{equation}
Spec(H(\widetilde{Q}^{a}))=Spec(H(Q^{a})),\text{ }\widetilde{I}%
(a)=I(a),\widetilde{I}_{1}(a,b)=I_{1}(a,\beta),\widetilde{I}_{2}%
(a,b)=I_{2}(a,b)
\end{equation}
\ (see Theorem 5 of [10]). We need \ to prove that $\widetilde{q}%
(x)\in\{q(sx+\tau):\tau\in F,s=\pm1\}.$ For this, it is enough to show that
there exist $\tau\in F,s\in\{-1,1\}$ such that $\widetilde{q}(sx-\tau)=q(x).$
The draft scheme of the proof is the followings. In Theorem 4 we proved that
if $q(x)$ has the form (3), then its Fourier coefficients $z(a)$ for $a\in
Q(1,1,1)$ can be defined uniquely, modulo (2), from the invariants (25)-(28),
(34) and (35). Here we prove that if the band functions of the operators
$L(q)$ and $L(\widetilde{q})$ coincide, then $\widetilde{q}$ has the form (3)
and the operator $L(\widetilde{q})$ has the spectral invariants, denoted by
($\widetilde{\text{25}}$)-($\widetilde{\text{28}}$), ($\widetilde{\text{34}}%
$), ($\widetilde{\text{35}}$) and obtained from the formulas (25)-(28), (34),
(35) respectively by replacing everywhere $z(a)$ with $\widetilde{z}(a).$
Then, using the arguments of the proof of Theorem 4 and fixing the inversion
and translations (2), we prove that $\widetilde{z}(a)=z(a)$ for $a\in
Q(1,1,1).$

Since $q^{a}(x)=0$ for $a\in S\backslash Q(1,1,1),$ the equality (15) and the
second equality of (106) imply that $\widetilde{q}$ has the form (93). Now, to
show that $\widetilde{q}(x)$ has the form (3), we prove that%
\begin{equation}
\widetilde{z}(na)=0,\text{ }\forall\mid n\mid>1,a\in Q(1,1,1).
\end{equation}
By (45) we have $Q^{\gamma_{k}}(s)=a_{k}\cos s$ where $a_{k}>0$ and $k=1,2,3.$
Therefore, by the first equality of (106), $\widetilde{Q}^{^{\gamma_{k}}}\in
M.$ On the other hand, by the definition of $W,$ we have $\widetilde
{Q}^{^{\gamma_{k}}}\in(C^{1}(\mathbb{R})\backslash M)\cup P$ (see (105). Thus
$\widetilde{Q}^{^{\gamma_{k}}}\in P.$ Then, it follows from Theorem 7 that
(107) holds for $a\in\{\gamma_{1},\gamma_{2},\gamma_{3}\}$. Hence the all
conditions of Lemma 2 hold and we have the formulas ($\widetilde{\text{25}}$),
($\widetilde{\text{26}}$) and ($\widetilde{\text{34}}$). Besides, it follows
from the second equality of (106) that$\mid\widetilde{z}(\gamma_{i})\mid=\mid
z(\gamma_{i})\mid.$ By Theorem 3 there exists $\tau\in F$ such that
\[
\arg(\widetilde{q}(x-\tau),e^{-i\langle\gamma_{k},x\rangle})=0,\text{ }\forall
k=1,2,3.
\]
Without loss of generality, we denote $\widetilde{q}(x-\tau)$ by
$\widetilde{q}$ and its Fourier coefficients by $\widetilde{z}(a).$ Thus we
have
\begin{equation}
\widetilde{z}(\gamma_{i})=z(\gamma_{i})=a_{i}>0,\text{ }\forall i=1,2,3.
\end{equation}
These with (25), (26), ($\widetilde{\text{25}}$), ($\widetilde{\text{26}}$)
and (108) imply that
\begin{equation}
\operatorname{Re}(\widetilde{z}(\gamma_{i}\pm\gamma_{j}))=\operatorname{Re}%
(z(\gamma_{i}\pm\gamma_{j})).
\end{equation}
From this using the obvious equalities (see (15) and the second equality of
(106))%
\begin{equation}%
%TCIMACRO{\tsum \limits_{n=1}^{\infty}}%
%BeginExpansion
{\textstyle\sum\limits_{n=1}^{\infty}}
%EndExpansion
2\mid\widetilde{z}(n(\gamma_{i}\pm\gamma_{j}))\mid^{2}=\widetilde{I}%
(\gamma_{i}\pm\gamma_{j})=I(\gamma_{i}\pm\gamma_{j})=2\mid z(\gamma_{i}%
\pm\gamma_{j})\mid^{2},
\end{equation}
we obtain
\begin{equation}
\mid\operatorname{Im}(\widetilde{z}(\gamma_{i}\pm\gamma_{j}))\mid\leq
\mid\operatorname{Im}(z(\gamma_{i}\pm\gamma_{j}))\mid.
\end{equation}
On the other hand, using (34), ($\widetilde{\text{34}}$), (108) and (106), we
obtain
\[
\operatorname{Re}(\widetilde{z}(\gamma_{i}+\gamma_{j})\widetilde{z}(\gamma
_{i}-\gamma_{j}))=\operatorname{Re}(z(\gamma_{i}+\gamma_{j})z(\gamma
_{i}-\gamma_{j})).
\]
This with (109) and (111) imply that
\begin{equation}
\mid\operatorname{Im}(\widetilde{z}(\gamma_{i}\pm\gamma_{j}))\mid
=\mid\operatorname{Im}(z(\gamma_{i}\pm\gamma_{j}))\mid.
\end{equation}
Thus by (109) and (112), we have
\begin{equation}
\mid\widetilde{z}(\gamma_{i}\pm\gamma_{j})\mid=\mid z(\gamma_{i}\pm\gamma
_{j})\mid.
\end{equation}
Therefore, from (110) we see that (107) holds for $a=\gamma_{i}\pm\gamma_{j}$.
Hence we have
\begin{equation}
\widetilde{z}(n(\gamma_{i}\pm\gamma_{j})=0,\text{ }\widetilde{z}(n\gamma
_{m})=0,\text{ }\forall n\in\mathbb{Z}\backslash\{-1,1\},
\end{equation}
where $i,j,m$ are different integers satisfying $1\leq i,j,m\leq3.$ Now
instead of (95) using (114), that is, instead $\gamma_{i}$ and $\gamma_{j}$ in
(95) taking $\gamma_{i}\pm\gamma_{j}$ and $\gamma_{m}$ respectively, and
repeating the proof of Lemma 2, we obtain that
\begin{equation}
\widetilde{I}_{1}(\gamma,\ \gamma_{i})=A_{1}(\gamma,\ \gamma_{i}%
)\operatorname{Re}(\widetilde{z}(-\gamma)\widetilde{z}(\gamma-\gamma
_{i})\widetilde{z}(\gamma_{i})), \tag{$\widetilde{27}$}%
\end{equation}%
\begin{equation}
\widetilde{I}_{1}(2\gamma_{i}-\gamma,\ \gamma_{i})=A_{1}(2\gamma_{i}%
-\gamma,\ \gamma_{i})\operatorname{Re}(\widetilde{z}(\gamma-2\gamma
_{i})\widetilde{z}(\gamma_{i}-\gamma)\widetilde{z}(\gamma_{i})),
\tag{$\widetilde{28}$}%
\end{equation}%
\begin{equation}
\widetilde{I}_{2}(\gamma_{i},\ \gamma-\gamma_{i})=A_{2}(\gamma_{i},\gamma
_{j})\operatorname{Re}(\widetilde{z}(-\gamma_{i}))^{2}\widetilde{z}%
(\gamma)\widetilde{z}(2\gamma_{i}-\gamma)) \tag{$\widetilde{35}$}%
\end{equation}
for $i=1,2,3;$ $i\neq j,$ where $\gamma=\gamma_{1}+\gamma_{2}+\gamma_{3}.$

One can readily see that the formulas ($\widetilde{\text{25}}$)-($\widetilde
{\text{28}}$), ($\widetilde{\text{34}}$), ($\widetilde{\text{35}}$) are
obtained from the formulas (25)-(28), (34), (35) respectively by replacing
everywhere $z(a)$ with $\widetilde{z}(a).$ Moreover, by (108), (109) and
(112), we have
\begin{equation}
\widetilde{a_{i}}=a_{i},\text{ }\forall i=1,2,...,6;\text{ }\widetilde{b_{i}%
}=\pm b_{i},\text{ }\forall i=4,5,6,
\end{equation}
where $\widetilde{a_{i}}+i\widetilde{b_{i}}=\widetilde{z}(\gamma_{i}).$ As in
\textit{Step 1} in the proof of Theorem 4, using ($\widetilde{\text{27}}$) for
$i=1,2,3$ and taking into account (115), we obtain the equations
\begin{equation}
a_{4}\widetilde{a}_{7}+\widetilde{t}_{4}\mid b_{4}\mid\widetilde{b}_{7}=c_{1}
\tag{$\widetilde{\text{50}}$}%
\end{equation}%
\begin{equation}
a_{5}\widetilde{a}_{7}+\widetilde{t}_{5}\mid b_{5}\mid\widetilde{b}_{7}=c_{2},
\tag{$\widetilde{\text{51}}$}%
\end{equation}%
\begin{equation}
a_{6}\widetilde{a}_{7}+\widetilde{t}_{6}\mid b_{6}\mid\widetilde{b}_{7}=c_{3},
\tag{$\widetilde{\text{52}}$}%
\end{equation}
where $\widetilde{t}_{m}$ is the sign of $\widetilde{b}_{m},$ that is, is
either $-1$ or $1$ and $c_{1},c_{2},c_{3}$ are the known constants defined in
(50), (51), (52). It follows from (46.1) that the main determinants of the
systems of equations, with respect to the unknowns $\widetilde{a}_{7},$
$\widetilde{b}_{7},$ generated by pairs \{($\widetilde{\text{50}}$),
($\widetilde{\text{51}}$)\}, \{($\widetilde{\text{50}}$), ($\widetilde
{\text{52}}$)\}, \{($\widetilde{\text{51}}$), ($\widetilde{\text{52}}$)\} are
not zero. Finding $\widetilde{b}_{7}$ from ($\widetilde{\text{50}}$),
($\widetilde{\text{51}}$) and taking into account (53), we see that
$\widetilde{b}_{7}\neq0.$ Therefore, for fixing the inversion $\widetilde
{q}(x)\longrightarrow\widetilde{q}(-x)$, we assume that $\widetilde{b}_{7}>0.$
Using this and finding $\widetilde{b}_{7}$ from the systems generated by pairs
\{($\widetilde{\text{50}}$), ($\widetilde{\text{51}}$)\}, \{($\widetilde
{\text{50}}$), ($\widetilde{\text{52}}$)\}, \{($\widetilde{\text{51}}$),
($\widetilde{\text{52}}$)\}, we get the inequalities
\begin{equation}
\frac{a_{4}c_{2}-a_{5}c_{1}}{\widetilde{t}_{5}\mid b_{5}\mid a_{4}%
-\widetilde{t}_{4}\mid b_{4}\mid a_{5}}>0,\text{ }\frac{a_{4}c_{3}-a_{6}c_{1}%
}{\widetilde{t}_{6}\mid b_{6}\mid a_{4}-\widetilde{t}_{4}\mid b_{4}\mid a_{6}%
}>0,\frac{a_{5}c_{3}-a_{6}c_{2}}{\widetilde{t}_{6}\mid b_{6}\mid
a_{5}-\widetilde{t}_{5}\mid b_{5}\mid a_{6}}>0 \tag{$\widetilde{\text{53}}$}%
\end{equation}
One can readily see that the relations ($\widetilde{\text{50}}$)-($\widetilde
{\text{53}}$) with respect to the unknowns $\widetilde{a}_{7},$ $\widetilde
{b}_{7},$ $\widetilde{t}_{4},$ $\widetilde{t}_{5}$, $\widetilde{t}_{6}$ are
obtained from (50)-(53) by replacing the unknowns $a_{7},$ $b_{7},$ $t_{4},$
$t_{5}$, $t_{6}$ with $\widetilde{a}_{7},$ $\widetilde{b}_{7},$ $\widetilde
{t}_{4},$ $\widetilde{t}_{5}$, $\widetilde{t}_{6}.$ Since we proved that
(50)-(53) has a unique solution, we have:

$a_{7}=\widetilde{a}_{7},$ $b_{7}=\widetilde{b}_{7},$ $t_{4}=\widetilde{t}%
_{4},$ $t_{5}=\widetilde{t}_{5},$ $t_{6}=\widetilde{t}_{6}$. This with (115)
imply that
\begin{equation}
\widetilde{a_{i}}=a_{i},\text{ }\widetilde{b_{i}}=b_{i},\text{ }\forall
i=1,2,...,7.
\end{equation}

In \textit{Step 2} and \textit{Step 3 }of Theorem 4 using the invariants (28),
(34), (35) we have determined the all other Fourier coefficients of $q$
provided that $a_{i}$ and $b_{i}$ for $i=1,2,...,7$ are known. Since the
invariants ($\widetilde{\text{28}}$), ($\widetilde{\text{34}}$),
($\widetilde{\text{35}}$) are obtained from the invariants (28), (34), (35) by
replacing everywhere $a_{i}$ and $b_{i}$ with $\widetilde{a_{i}}$ and
$\widetilde{b_{i}}$ respectively, and (116) holds, we have
\begin{equation}
\widetilde{z}(a)=z(a),\text{ }\forall a\in Q(1,1,1).
\end{equation}
This with the equalities (15), (20), (106) and (94) imply that (107) holds for
all $a\in Q(1,1,1)$. Therefore, it follow from (93), (107) and (117) that
$\widetilde{q}(x)=q(x)$
\end{proof}

\begin{acknowledgement}
The work was supported by the Scientific and Technological Research Council of
Turkey (T\"{u}bitak, project No. 108T683).
\end{acknowledgement}

\end{document}